\documentclass[paper]{ieice}
\usepackage{graphicx,xcolor}
\usepackage[fleqn]{amsmath}
\usepackage{newtxtext}
\usepackage[varg]{newtxmath}
\usepackage{cite}
\usepackage{amsmath}
\usepackage{bm}
\usepackage{algorithm}
\usepackage{algorithmic}
\usepackage{booktabs}
\usepackage{physics}
\DeclareMathOperator{\sinc}{sinc}

\usepackage{graphicx}
\usepackage{textcomp}
\usepackage{xcolor}
\usepackage{verbatim}

\field{}
\title{}

\received{2026}{2}{23}
\revised{2026}{7}{??}

\begin{document}
\title{A Parallel Two-Stage Prony Framework for Robust Delay-Doppler Estimation
}
\titlenote{A part of this paper has been presented in PIMRC 2025 and ICASSP 2026. This work was supported in part by JSPS KAKENHI Number JP23K26104 and JP23H00474.}

\authorlist{%
 \authorentry[sun@me.inf.kyushu-u.ac.jp]{Liangchen Sun}{n}{labelA}\MembershipNumber{}
 \authorentry[jitsumatsu@inf.kyushu-u.ac.jp]{Yutaka Jitsumatsu}{m}{labelB}\MembershipNumber{9907289}
}

\affiliate[labelA]{%
Department of Information Science and Technology, Kyushu University, Japan
}
\affiliate[labelB]{%
Department of Informatics, Kyushu University, Japan
}

\maketitle

\begin{summary}  
This paper addresses off-grid physical delay–Doppler parameter estimation based on an orthogonal time frequency space (OTFS) pilot waveform over doubly selective channels. Based on a known single-pilot waveform with a frequency-shifted sinc pulse structure, we propose a parallel dual-order estimation framework, in which Prony-based Doppler-first and delay-first procedures are executed concurrently and their candidate paths are jointly fused. The two estimation orders provide complementary resolution characteristics, thereby mitigating ambiguities caused by closely spaced multipath components and fractional delay--Doppler mismatch. Numerical results show that the proposed framework improves detection reliability and estimation stability compared with a single-order Prony estimator, especially in off-grid scenarios and low-to-moderate signal-to-noise ratio regimes.

\end{summary}


\section{Introduction}

In high-mobility wireless environments, multipath propagation combined with user mobility gives rise to doubly selective channels, which are characterized by simultaneous delay and Doppler spreads~\cite{Delay-Doppler-Communications}. Under such conditions, the loss of signal orthogonality caused by time variation and frequency dispersion leads to severe performance degradation, making accurate delay--Doppler parameter estimation and reliable sensing fundamentally challenging. Although orthogonal frequency division multiplexing (OFDM) has been widely adopted in modern wireless systems, its performance deteriorates in the presence of significant Doppler spread due to inter-carrier interference (ICI). Basis expansion model (BEM)-based approaches can approximate time-varying channels using predefined basis functions~\cite{Hlawatsch-Book}, but they generally require a large number of parameters and may incur high computational complexity, which limits their suitability for high-mobility and low-latency scenarios.

To overcome the limitations of conventional OFDM in high-mobility scenarios, delay--Doppler-domain waveforms have attracted considerable attention. Among them, orthogonal time-frequency space (OTFS) modulation is a representative technique that maps information symbols onto the delay--Doppler domain and provides an approximately time-invariant two-dimensional representation of rapidly time-varying channels~\cite{OTFS, OTFS-BITS, mohammed2023otfs}. This representation enables multipath components to be characterized by their delay and Doppler parameters and improves robustness against mobility-induced channel variations compared with conventional OFDM~\cite{Raviteja2018}. These properties also make OTFS attractive for high-mobility sensing and integrated sensing and communication (ISAC), where accurate delay--Doppler parameter estimation is essential.

Despite these advantages, accurate estimation of physical delay and Doppler parameters remains a bottleneck. Many existing implementations assume that channel paths or targets are located on discrete delay--Doppler grid points. In practice, however, physical propagation parameters are continuous and generally do not align with predefined grids. This mismatch leads to energy leakage across neighboring bins, significantly degrading the performance of conventional grid-based estimation methods and adversely affecting target/path detection, delay--Doppler map construction, and subsequent sensing or channel-aware processing~\cite{zegrar2024otfs,zhang2023radar,zacharia2023fractional,li2022novel,muppaneni2023channel,li2024grid,wang20222d,sheng2023time}. 

Consequently, off-grid delay–Doppler parameter estimation has become a critical challenge in fully exploiting the potential of OTFS for sensing-oriented applications. To address this issue, several solutions have been proposed, which can be broadly categorized into three classes. The first class relies on grid refinement combined with sparse recovery, including OMP-based parametric estimation~\cite{zegrar2024fractional,zacharia2023fractional} and sparse Bayesian learning (SBL)~\cite{zhang2024sparse,li2025fractional}. Although these approaches can improve estimation accuracy, they usually incur high computational complexity and exhibit sensitivity to grid resolution. The second class explicitly models fractional-induced inter-path interference and employs iterative cancellation strategies~\cite{muppaneni2023channel}; however, their performance is often sensitive to noise levels and path ordering, especially in dense multipath environments. The third class focuses on low-complexity designs based on integer delay--Doppler assumptions or separability properties~\cite{khan2023low,yogesh2024low}, but suffers from significant performance degradation under severe off-grid mismatch. More recently, we proposed a two-stage Prony-based delay--Doppler estimation method~\cite{YutakaPIMRC2025}, which achieves high accuracy at high signal-to-noise ratio (SNR) but exhibits reduced robustness in low-SNR or dense multipath scenarios. Overall, existing methods either suffer from grid dependence and computational burden or lack robustness, motivating the development of more reliable and efficient off-grid physical parameter estimation frameworks.

Motivated by these developments, this paper proposes a parallel dual-order Prony-based delay–Doppler parameter estimation framework based on an OTFS pilot waveform.
The adopted pilot waveform leads to a parametric sampled signal model, from which Doppler-first and delay-first estimation procedures are executed in parallel to exploit the complementary resolution characteristics of the two estimation orders.
The resulting parameter candidates are then fused through a consistency-driven mechanism, which suppresses spurious estimates and improves detection reliability and estimation stability without introducing substantial computational overhead.
The proposed framework is particularly suitable for high-mobility sensing and ISAC scenarios where robust off-grid physical delay--Doppler parameter estimation is required.

\section{Problem Formulation}

This section presents the OTFS pilot waveform used for sensing-oriented off-grid delay--Doppler parameter estimation, the corresponding doubly selective physical channel model, and the discrete-time observations employed by the proposed estimator\footnote{\ref{sec:model} compares the realization adopted in this work with three other DD-domain waveform realizations.}. To mitigate boundary effects, two additional guard components are introduced in both the time and frequency domains. Accordingly, the transmitted pilot waveform is generated using \(N+2\) pulses and \(M+2\) frequency lines, while the original \(N\times M\) DD grid is retained for parameter estimation.

\subsection{Transmitted Signal}
\label{sect:transmitted_signal}

We consider an OTFS waveform with \(M\) delay-domain samples and \(N\) Doppler-domain samples per transmission frame. Let \(T\) denote the delay-domain period, which
corresponds to a frequency-line spacing of \(1/T\) in the Dirichlet representation. Consequently, the nominal DD frame duration is \(NT\), the bandwidth of the central \(M\) frequency lines is \(M/T\), the delay resolution is \(T/M\), and the Doppler resolution is \(1/(NT)\). The delay--Doppler (DD)-domain symbols are denoted by \(X_{\mathrm{DD}}[k,\ell]\), where \(k \in \{0,\ldots,N-1\}\) and \(\ell \in \{0,\ldots,M-1\}\). Throughout this work, the DD-domain signal is assumed to be periodic along both dimensions, i.e.,
\(X_{\mathrm{DD}}[k+N,\ell]=X_{\mathrm{DD}}[k,\ell]\) and
\(X_{\mathrm{DD}}[k,\ell+M]=X_{\mathrm{DD}}[k,\ell]\).
For the realization adopted in this work, the time-domain samples are obtained by applying the inverse discrete Zak transform (IDZT)~\cite{yogesh2024low,lampel2022discrete} to the DD-domain symbols:
\begin{align}
x_{\mathrm{TD}}[nM+\ell]
=
\frac{1}{N}
\sum_{k=0}^{N-1}
X_{\mathrm{DD}}[k,\ell]\,
W_N^{-nk},
\label{eq:idzt}
\end{align}
where \(W_N=e^{-j2\pi/N}\) denotes the twiddle factor of the \(N\)-point discrete Fourier transform (DFT).
Based on these samples, the transmitted complex baseband signal is expressed as
\begin{align}
s(t)
=
\sum_{n=0}^{N-1}
\sum_{\ell=0}^{M-1}
x_{\mathrm{TD}}[nM+\ell]\,
p_{\mathrm{T}}\!\left(
t-(nM+\ell)\frac{T}{M}
\right),
\label{eq:transmitted_signal}
\end{align}
where \(p_{\mathrm{T}}(t)\) denotes the elementary transmit pulse used for continuous-time pulse shaping.

For sensing-oriented delay--Doppler parameter estimation, we consider a single nonzero pilot located at the origin of the DD grid.
This pilot configuration enables direct observation of the delay--Doppler response induced by physical propagation paths.
Specifically, the DD-domain symbol is defined as
\(X_{\mathrm{DD}}[k,\ell]=1\) for \(k=\ell=0\), and
\(X_{\mathrm{DD}}[k,\ell]=0\) for all other index pairs.
Under this pilot configuration, the corresponding time-domain samples reduce to
\begin{align}
x_{\mathrm{TD}}[nM+\ell]
=
\begin{cases}
\frac{1}{N}, & \ell=0,\; n=0,1,\ldots,N-1,\\
0, & \text{otherwise}.
\end{cases}
\label{pilot-DDdomain}
\end{align}
The constant normalization factor does not affect the estimation of delay and Doppler parameters and can be absorbed into the path gains.

For the central \(M\)-line component of the continuous-time realization used in this work, the elementary transmit pulse \(p_{\mathrm{T}}(t)\) is chosen as a frequency-shifted ideal sinc pulse rather than a rectangular transmit pulse~\cite{raviteja2019orthogonal,zhang2021modulation,keskin2021radar,gaudio2020effectiveness,shi2023integrated,ranasinghe2024fast}. Specifically, for even \(M\), its Fourier transform is defined as
\begin{align}
P_{\mathrm{T}}(f)
=
\begin{cases}
1, & -\dfrac{M+1}{2T} \le f < \dfrac{M-1}{2T}, \\[4pt]
0, & \text{otherwise},
\end{cases}
\label{eq:pulse_frequency}
\end{align}
where the asymmetric spectral support results from an intentional frequency shift. The \(M\) discrete frequency bins are indexed as \(-M/2,-M/2+1,\ldots,M/2-1\), whose center frequency is \(-1/(2T)\). Accordingly, the ideal rectangular spectrum is shifted by \(-1/(2T)\) so that its spectral support is centered with respect to these \(M\) discrete frequency bins. The corresponding time-domain transmit pulse is
\begin{align}
p_{\mathrm{T}}(t)
=
\frac{\sin\!\left(\pi M t/T\right)}{\pi t}
e^{-j\pi t/T}.
\label{eq:pulse_time}
\end{align}

With the single-pilot configuration in~\eqref{pilot-DDdomain}, substituting the time-domain samples into~\eqref{eq:transmitted_signal} yields a pilot waveform formed by shifted copies of the frequency-shifted sinc transmit pulse. Over the retained interval exploited by the estimator, we adopt a periodic approximation of the finite sinc pulse train. The corresponding periodic waveform admits the following Dirichlet-type finite sum-of-exponentials representation:
\begin{align}
\frac{1}{T}
\sum_{m=-M/2}^{M/2-1}
e^{j m \frac{2\pi}{T} t}
=
\frac{\sin\!\left(M\pi t/T\right)}
{T\sin\!\left(\pi t/T\right)}
e^{-j\pi t/T}.
\label{dirichlet_waveform}
\end{align}

\begin{figure} [t]
    \centering 
    \includegraphics[width=0.65\linewidth]{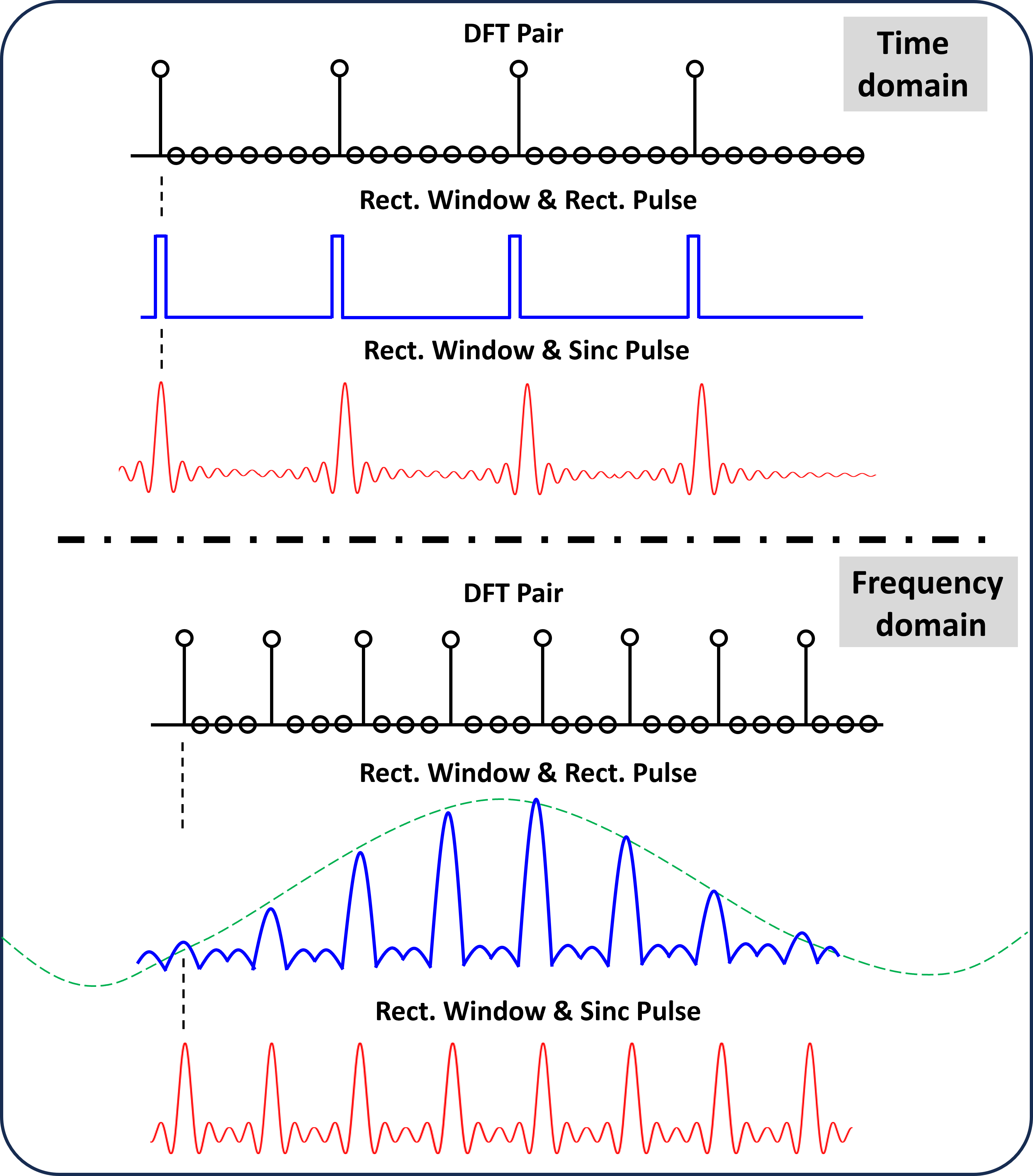} 
    \caption{Comparison of continuous-time waveform reconstructions obtained from rectangular and sinc transmit pulses for the same discrete-time samples, shown in the time domain (upper) and frequency domain (lower), with \(N=4\) and \(M=8\).}
    \label{fig:dirichlet_vs_sinc_pulses} 
\end{figure}

The key reason for adopting this pulse is that the resulting pilot waveform exhibits an approximately periodic structure and admits a finite sum-of-exponentials representation. After the received samples are arranged into time-domain or frequency-domain matrices, the delay and Doppler parameters appear as exponential modes in separable matrix factors, which is the key structure exploited by the proposed two-stage Prony estimator. In contrast, a rectangular transmit pulse does not yield the finite sum-of-exponentials structure required by the proposed Prony formulation, as illustrated in Fig.~\ref{fig:dirichlet_vs_sinc_pulses}.

However, Fig.~\ref{fig:dirichlet_vs_sinc_pulses} also shows that the adopted pilot waveform exhibits sinc-like sidelobes around the dominant pulse locations in the time domain. In the present pilot-based sensing setting, these deterministic sidelobes do not constitute unknown inter-symbol interference. The transmitted pilot waveform \(s(t)\) is known at the receiver, and each propagation path is modeled by a delayed and Doppler-shifted replica \(s(t-t_{d,p})e^{j2\pi f_{D,p}t}\). Hence, the sidelobes are part of the known waveform model rather than unmodeled interference. In practical data transmission systems, additional guard intervals, cyclic extension, or receiver filtering/windowing may be required to control inter-symbol interference, but such waveform-level data-transmission issues are outside the scope of the present physical parameter estimation study.

\begin{figure}[t] 
    \centering 
    \includegraphics[width=0.8\linewidth, height=0.45\textheight, keepaspectratio]{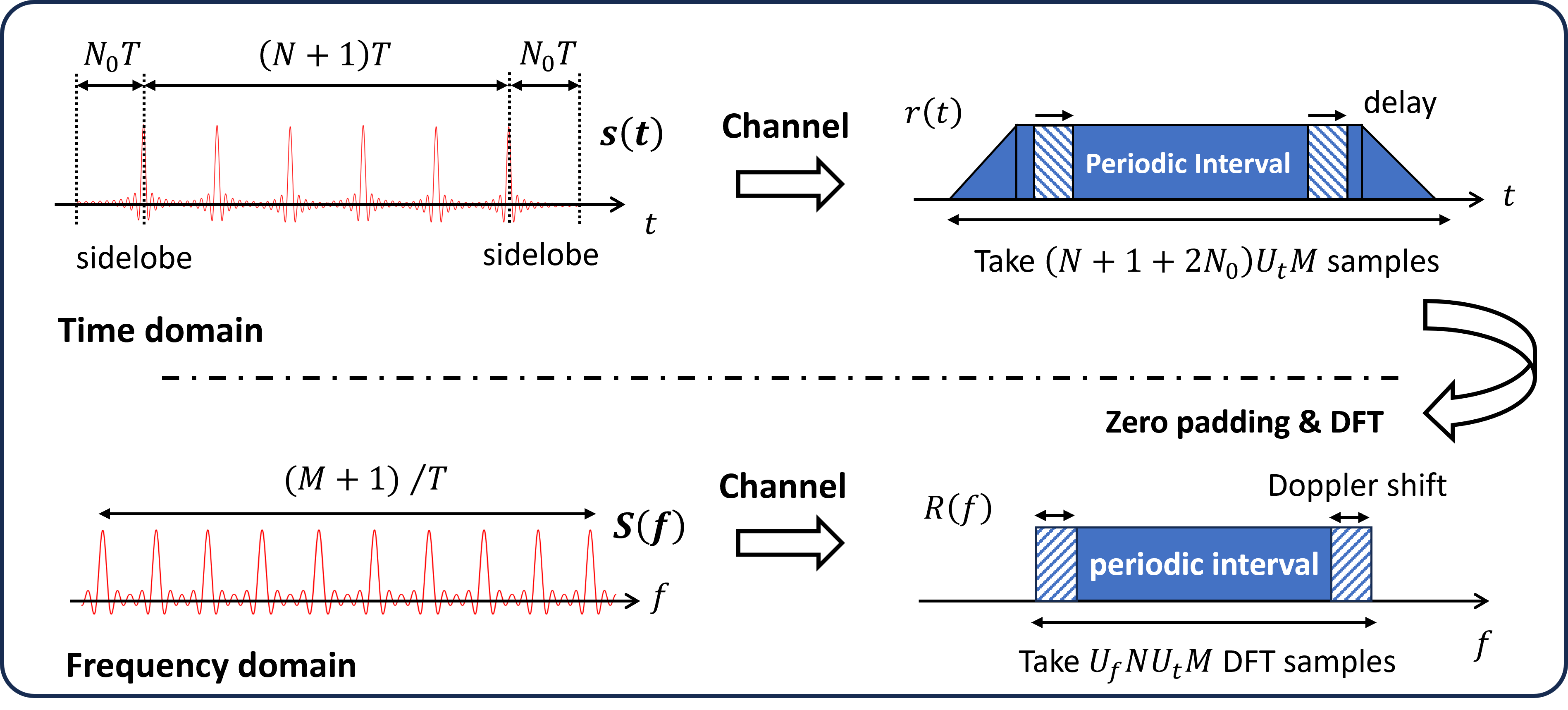}
    \capwidth=1\linewidth 
    \caption{Time-domain and frequency-domain samples of the adopted OTFS pilot waveform.}
    \label{Fig.2} 
\end{figure}

\subsection{Received Signal}
\label{sect:channel_model}

Based on~(\ref{eq:transmitted_signal}), we consider a multipath doubly selective physical channel and derive the sampled observation model used by the proposed estimator.
The number of propagation paths is denoted by \(P\).
The value of \(P\) is unknown at the receiver, but the channel is assumed to be sparse, i.e., \(P \ll NM\), where \(NM\) represents the effective dimension of the signal space.
Let \(t_{d,\max}\) and \(f_{D,\max}\) denote the maximum delay and the maximum Doppler shift, respectively.
For the \(p\)-th propagation path, \(\alpha_p\), \(t_{d,p}\), and \(f_{D,p}\) denote the complex attenuation coefficient, time delay, and Doppler shift, respectively.
The attenuation coefficients \(\alpha_p\) are modeled as independent complex Gaussian random variables.
The time-slot duration \(T\) is chosen such that \(0<t_{d,p}<T\) and \(-1/(2T)<f_{D,p}<1/(2T)\) for all \(p\).
Equivalently, \(T\) is selected to cover the maximum path delay while avoiding Doppler ambiguity within the considered normalized Doppler range.
Under these assumptions, the continuous-time received waveform is expressed as
\begin{align}
r(t)
=
\sum_{p=1}^{P}
\alpha_p\, s(t-t_{d,p})\, e^{j2\pi f_{D,p}t}
+ z(t),
\label{r(t)}
\end{align}
where \(z(t)\) denotes additive white Gaussian noise (AWGN).
The path-dependent constant phase term associated with the delay-Doppler channel convention is absorbed into the complex gain \(\alpha_p\).

Due to the presence of delay and Doppler shifts, the finite periodicity of the transmitted pilot waveform is preserved only over a reduced interval compared with the ideal unshifted case. In particular, boundary regions may be affected by delayed waveform tails and Doppler-induced spectral spreading. To mitigate these boundary effects, as illustrated in Fig.~\ref{Fig.2}, the transmitted pilot waveform used in the simulations is generated with additional guard components in both the time and frequency domains. Specifically, the upper summation limit in \eqref{eq:transmitted_signal} is extended from \(N-1\) to \(N+1\), resulting in \(N+2\) time-domain pulses with a nominal transmit duration of \((N+2)T\). Similarly, the \(M\) central frequency lines described in \eqref{eq:pulse_frequency}--\eqref{dirichlet_waveform} are supplemented by one guard line on each side, at indices \(m=-M/2-1\) and \(m=M/2\). Thus, the transmitted waveform used in the simulations contains \(M+2\) frequency lines with an occupied bandwidth of \((M+2)/T\). These guard components are discarded from the corresponding observations before Prony estimation, as described in the following sections. Therefore, for clarity, the main-text formulations retain the original \(N\) time intervals and \(M\) central frequency lines used for parameter estimation, while the guard extensions are specified separately.

The receiver front-end is modeled as an ideal low-pass filter followed by uniform sampling. Let \(r_{\mathrm{L}}(t)\) denote the low-pass-filter output. The sampling interval is \(T_s=T/(U_tM)\), where \(U_t=2\) is the time-domain oversampling factor used throughout this work. Under the ideal LPF model described in~\ref{app:sampling}, the useful-signal component passes through the filter without distortion, and the filtered samples coincide exactly with the corresponding continuous-time point samples. Accordingly,
\begin{align}
r_{\mathrm{TD}}[i]
=
\sum_{p=1}^{P}
\alpha_p
s(iT_s-t_{d,p})
e^{j2\pi f_{D,p}iT_s}
+
z_{\mathrm{TD}}[i],
\label{eq:rtd_sampled}
\end{align}
where \(z_{\mathrm{TD}}[i]\) denotes the receive-filtered and sampled noise component, which is white with variance \(N_0/T_s\). The sampling model and its justification are provided in~\ref{app:sampling}. Equation~\eqref{eq:rtd_sampled} is used as the sampled pilot observation model for estimating the continuous physical delay and Doppler parameters from the known OTFS pilot waveform.

Let \(R(f)\) denote the Fourier transform of the low-pass-filtered received waveform \(r_{\mathrm{L}}(t)\).
Its frequency-domain samples are defined as \(R_{\mathrm{FD}}[k]=R(k\Delta f_{\mathrm{FD}})\), where \(\Delta f_{\mathrm{FD}}=1/(U_fNT)\) denotes the frequency-domain sampling interval and \(U_f=2\) is the frequency-domain oversampling factor used throughout this work. The frequency-domain samples are computed as
\begin{align}
R_{\mathrm{FD}}[k]
&=
\int r_{\mathrm{L}}(t)\,
e^{-j 2\pi \frac{k}{U_f N T} t}\, dt
\notag\\
&\approx
T_s \sum_{i}
r_{\mathrm{L}}(i T_s)\,
e^{-j 2\pi \frac{k}{U_f N T} i T_s}
\notag\\
&=
T_s \sum_{i}
r_{\mathrm{TD}}[i]\,
e^{-j 2\pi \frac{k i}{U_t U_f N M}},
\label{Rfd}
\end{align}
where the second line is a discrete approximation of the Fourier integral over the retained observation interval. Its accuracy improves with the time-domain oversampling factor and with a sufficiently large observation margin. The summation range of \(i\) is selected to capture the significant portions of the sum-of-sinc pilot waveform in the time domain.

Since the adopted frequency-shifted sinc pilot waveform is not strictly time-limited, its sidelobes may extend beyond the nominal observation interval after delay and Doppler shifts.
To reduce boundary truncation mismatch and preserve the effective finite-periodic structure required for the subsequent DFT-based processing, an observation margin of length \(N_0T\) is retained on both sides of the transmitted pilot waveform.
Accordingly, in addition to the \(N\) dominant peaks of the pilot waveform, two additional peaks are included, whose centers are located at \(t=0,T,\ldots,(N+1)T\).
With the retained tail regions, the total temporal observation interval becomes \([-N_0T,\,(N+1+N_0)T]\), resulting in \((N+1+2N_0)U_tM\) discrete-time samples.
To obtain the DFT coefficients used in the delay-first procedure, the sample sequence is further zero-padded to a length of \(U_fU_tNM\).

\section{Proposed Method}
\label{sect:proposed_method}

This section presents the proposed parallel delay--Doppler estimation framework based on the sampled OTFS pilot observation model developed in Section~\ref{sect:channel_model}. The framework consists of two complementary branches: a Doppler-first branch and a delay-first branch. The Doppler-first branch first estimates Doppler candidates from the time-domain samples and then recovers the associated delays, whereas the delay-first branch performs the estimation in the reverse order using frequency-domain samples. Both branches use fixed candidate orders, namely \(\hat P_D=N-1\) and \(\hat P_d=M-1\). Their candidates are subsequently fused and pruned to determine the physically valid delay--Doppler paths and the final number of detected components.

\subsection{Doppler-first Procedure}
\label{sect:primal}

We first describe the Doppler-first estimation procedure. For notational simplicity, the following derivation isolates the noiseless separable structure exploited by the estimator. In implementation, the estimator is applied directly to noisy observations, and the required coefficients and path gains are obtained using least-squares and pseudoinverse operations.

Due to the presence of path delays, the finite periodicity of \(r_{\mathrm{TD}}[i]\) is broken near the temporal boundaries. Accordingly, the two temporal guard regions introduced in Section~\ref{sect:channel_model} are discarded before Prony estimation. The retained samples correspond to the half-open interval \(t\in[T,(N+1)T)\), which contains exactly \(NU_tM\) samples and preserves the finite-periodic structure required for Prony-based estimation. After shifting the time origin by \(T\), these samples are relabeled as \(i=0,1,\ldots,NU_tM-1\). This shift introduces only a path-dependent constant Doppler phase, which can be absorbed into the complex path gain \(\alpha_p\), and therefore does not change the delay or Doppler parameters.

From~\eqref{eq:rtd_sampled}, the noiseless component of the received time-domain samples over the preserved interval can be modeled as
\begin{equation}
r_{\mathrm{TD}}[i]
=
\sum_{p=1}^{P}
\alpha_p
s(iT_s-t_{d,p})
e^{j2\pi f_{D,p}iT_s},
\label{eq:rTD_periodic}
\end{equation}
where \(i=0,1,\ldots,NU_tM-1\).
To reveal the inherent structure of the received signal, we reorganize the samples into an \(N\times(U_tM)\) matrix \(\bm R\) by introducing the index decomposition $i=nU_tM+\ell,$ and defining
$R_{n,\ell}\triangleq r_{\mathrm{TD}}[nU_tM+\ell],$ where \(n=0,1,\ldots,N-1\) and \(\ell=0,1,\ldots,U_tM-1\). 
Using $iT_s=nT+\ell T_s$ and the periodic approximation adopted over the retained interval, the \((n,\ell)\)-th entry of \(\bm R\) is given by
\begin{align}
R_{n,\ell}
&=
\sum_{p=1}^{P}
\underbrace{
e^{j2\pi f_{D,p}nT}
}_{E_{n,p}}
\nonumber\\
&\quad\times
\underbrace{
\alpha_p s(\ell T_s-t_{d,p})
e^{j2\pi f_{D,p}\ell T_s}
}_{V_{p,\ell}} ,
\label{eq:R_separable_underbrace}
\end{align}
which reveals a bilinear separable structure across the Doppler and delay dimensions. 
The Doppler-dependent terms form a Vandermonde matrix, while the delay-dependent information is embedded in the remaining factor. 
Accordingly, we define
\begin{equation}
E_{n,p} \triangleq e^{j2\pi f_{D,p}nT}.
\label{eq:E_def}
\end{equation}
\begin{equation}
V_{p,\ell} \triangleq
\alpha_p s(\ell T_s-t_{d,p})
e^{j2\pi f_{D,p}\ell T_s}.
\label{eq:V_def}
\end{equation}
Then, \eqref{eq:R_separable_underbrace} can be compactly written as
\begin{equation}
R_{n,\ell}
=
\sum_{p=1}^{P}E_{n,p}V_{p,\ell},
\end{equation}
which leads to the matrix representation
\begin{equation}
\bm R=\bm E\bm V.
\label{eq:R_EV}
\end{equation}

The factorization in~\eqref{eq:R_EV} shows that the Doppler shifts \(\{f_{D,p}\}\) determine the Vandermonde structure of \(\bm E\), whereas the delay parameters \(\{t_{d,p}\}\) are included in \(\bm V\). 
This property motivates a sequential estimation strategy: Doppler frequencies are first recovered, and the corresponding delays are then estimated from the Doppler-compensated measurements.
Once the Doppler frequencies are estimated, we compensate for the Doppler-induced phase rotations in $\bm V$,
yielding
\begin{align}
    \widetilde V_{p,\ell} = \alpha_p s(\ell T_s - t_{d,p}).
\end{align}

To extract delay information from $\widetilde V_{p,\ell}$, we exploit the Dirichlet representation of the transmitted waveform.
Taking the $U_tM$-point DFT of $\widetilde V_{p,\ell}$ over $\ell=0,1,\ldots,U_tM-1$ yields
\begin{align}
&\sum_{\ell=0}^{U_tM-1}
\widetilde V_{p,\ell}
e^{-j\frac{2\pi}{U_tM}m\ell}
\nonumber\\
&=
\sum_{\ell=0}^{U_tM-1}
\alpha_p s(\ell T_s-t_{d,p})
e^{-j\frac{2\pi}{U_tM}m\ell}
\nonumber\\
&\stackrel{\mathrm{(a)}}{=}
\alpha_p
\sum_{\ell=0}^{U_tM-1}
\sum_{m'=-M/2}^{M/2-1}
e^{jm'\frac{2\pi}{T}(\ell T_s-t_{d,p})}
e^{-j\frac{2\pi}{U_tM}m\ell}
\nonumber\\
&=
\alpha_p
\sum_{m'=-M/2}^{M/2-1}
e^{-j2\pi m't_{d,p}/T}
U_tM
\delta_{m'-m\ {\rm mod}\ U_tM}
\nonumber\\
&=
\left\{
\begin{array}{l}
\alpha_p U_tM e^{-j2\pi m t_{d,p}/T},\\[1mm]
\qquad\text{if }-\dfrac{M}{2}\le m\le\dfrac{M}{2}-1
\ {\rm mod}\ U_tM,\\[2mm]
0,\qquad\text{otherwise}.
\end{array}
\right.
\label{DFT_of_V_tilde}
\end{align}
where step (a) follows from substituting the Dirichlet expansion in~\eqref{dirichlet_waveform}. 
Known constant scaling factors are omitted since they can be absorbed into the path gains and do not affect the estimation of delay or Doppler parameters.

Equations~\eqref{eq:E_def} and~\eqref{DFT_of_V_tilde} show that both \(f_{D,p}\) and \(t_{d,p}\) appear in complex-exponential form and therefore admit a standard Prony formulation. Consequently, they can be efficiently retrieved by applying Prony's method to the corresponding exponential-mode observations.

\subsubsection{Stage 1: Doppler Estimation}

In the Doppler-first branch, the first stage recovers Doppler candidates associated with the propagation paths. 
Although \(\bm R\) contains \(U_tM\) columns, each column shares the same Doppler-dependent Vandermonde structure. 
Rather than applying Prony's method to a single observation vector, the proposed method exploits all \(U_tM\) columns jointly, which improves robustness against noise.

To estimate the Doppler modes, we construct an auxiliary matrix \(\bm T\) by reversing the row order of \(\bm R\) and taking its transpose, i.e., \(T_{\ell,n}=R_{N-1-n,\ell}\), where \(n=0,1,\ldots,N-1\) and \(\ell=0,1,\ldots,U_tM-1\). Let $\bm a_D=
\bigl(a_D[0],a_D[1],\ldots,a_D[\hat P_D]\bigr)^{\mathrm T},$ with $a_D[0]=1.$ Under noisy observations, the annihilation relation is imposed in the least-squares sense:
\begin{equation}
\bm a_D
=
\arg\min_{\bm a:\,a[0]=1}
\left\|
\bm T\bm a
\right\|_2^2.
\end{equation}
Let \(\bm t_0\) denote the first column of \(\bm T\), and let \(\widetilde{\bm T}\) denote the submatrix obtained by removing this column. Writing $\bm a_D^{\mathrm T}
= \bigl( 1,\widetilde{\bm a}_D^{\mathrm T} \bigr),$ the remaining coefficients are obtained as $\widetilde{\bm a}_D = -\widetilde{\bm T}^{\dagger}\bm t_0,$ where \((\cdot)^\dagger\) denotes the Moore--Penrose pseudoinverse. The Doppler-related modes are then obtained as the roots of
\begin{align}
a_D[0]x^{\hat P_D}
+
a_D[1]x^{\hat P_D-1}
+\cdots+
a_D[\hat P_D]
=
0,
\label{polynomial}
\end{align}
which are denoted by
\(\{Z_{D,p}\}_{p=1}^{\hat P_D}\).
The Doppler estimates are finally obtained from their phases as 
\begin{align}
\hat{f}_{D,p} = \frac{\arg(Z_{D,p})}{2\pi T},
\end{align}
where $\arg(\cdot)$ denotes the principal value of the complex argument, taking values in $[-\pi,\pi]$.

\subsubsection{Doppler Compensation and Delay Extraction}

Given the obtained Doppler candidates \(\{\hat f_{D,p}\}_{p=1}^{\hat P_D}\), we  reconstruct the Doppler matrix as $\widehat{\bm E}=\left(e^{j2\pi \hat f_{D,p}nT}\right)_{n,p}.$
Substituting \(\widehat{\bm E}\) into~\eqref{eq:R_EV}, the coefficient matrix \(\bm V\) is estimated by least squares:
\begin{align}
\widehat{\bm V}
&=
\arg\min_{\bm V}
\left\|
\bm R-\widehat{\bm E}\bm V
\right\|_F^2
=
\widehat{\bm E}^{\dagger}\bm R,
\label{eq:V_hat_LS}
\end{align}

To extract the delay-dependent structure, the Doppler-induced phase term in \(\widehat{\bm V}\) is compensated as
\begin{align}
\widetilde V_{p,\ell}
=
\widehat V_{p,\ell}
e^{-j2\pi \hat f_{D,p}\ell T_s}.
\label{eq:V_tilde_comp}
\end{align}
For each candidate index \(p\), a \(U_tM\)-point DFT is then applied to \(\widetilde V_{p,\ell}\), yielding
\begin{align}
Y_p[m]
=
\sum_{\ell=0}^{U_tM-1}
\widetilde V_{p,\ell}
e^{-j\frac{2\pi}{U_tM}m\ell}.
\label{eq:Yp_def}
\end{align}

In the ideal case where the Doppler estimates are exact and noise is absent, \eqref{DFT_of_V_tilde} implies that
\begin{align}
Y_p[m]
=
\alpha_p U_tM
e^{-j2\pi m t_{d,p}/T}
\label{eq:Yp_ideal}
\end{align}
over the central frequency-index interval $-\frac{M}{2}\le m\le\frac{M}{2}-1$ mod $\ U_tM$.
Thus, \(Y_p[m]\) contains the delay \(t_{d,p}\) as the phase of a complex exponential. 
Only the central \(M\) frequency bins are retained for the subsequent Prony estimation, while the two additional guard lines introduced during waveform generation are discarded.

\subsubsection{Stage 2: Delay Estimation}

Having obtained the Doppler candidates in Stage~1, the second stage recovers their associated delay candidates. In our previous study~\cite{YutakaPIMRC2025}, the estimator allowed multiple propagation paths to share the same Doppler component, where \(L\) denotes the number of delay modes associated with each Doppler candidate. However, when Doppler frequencies coincide or are closely spaced, separating such paths using the Doppler-first branch alone becomes difficult.

In the present parallel framework, we set \(L=1\) in the Doppler-first branch and rely on the delay-first branch to provide complementary candidates for paths with close Doppler shifts. This choice also reduces the computational cost of the second stage. For each Doppler-compensated spectrum $\bm Y_p= \bigl(Y_p[m]\bigr)_{m=-M/2}^{M/2-1},$
we construct the \((M-1)\times2\) Toeplitz matrix $(\bm T_p)_{ij} = Y_p\!\left[ 1-\frac{M}{2}+i-j \right],$ where \(i=1,\ldots,M-1\) and \(j=1,2\). The coefficient vector $\bm a_d= \bigl(a_d[0],a_d[1]\bigr)^{\mathrm T}$ is estimated from the approximate null space of \(\bm T_p\) and normalized such that \(a_d[0]=1\). The root of the resulting first-order polynomial is denoted by \(Z_{d,p}\).
The corresponding delay for the Doppler estimate $\hat f_{D,p}$ is then recovered as
\begin{align}
    \hat{t}_{d,p} = T \left( -\frac{\arg(Z_{d,p})}{2\pi} \bmod 1 \right),
    \label{hat_t}
\end{align}
where the modulo operation ensures that $\hat{t}_{d,p}$ lies within $[0, T)$.

\subsection{Delay-first Procedure}
\label{sect:dual}

Next, we describe the delay-first procedure, which follows a framework analogous to the Doppler-first procedure, with the roles of delay and Doppler interchanged. 
The delay-first branch exploits frequency-domain samples of the received pilot waveform and provides complementary estimates to the Doppler-first branch. According to~\eqref{Rfd}, the frequency-domain samples \(R_{\mathrm{FD}}[k]\) can be expressed as
\begin{equation}
R_{\mathrm{FD}}[k]
=
\sum_{p=1}^{P}
\alpha_p
S\!\left(
\frac{k}{U_fNT}-f_{D,p}
\right)
e^{-j2\pi t_{d,p}\frac{k}{U_fNT}},
\label{eq:RFD_recall}
\end{equation}
where \(S(f)\) denotes the Fourier transform of \(s(t)\).
As in the Doppler-first branch, the following derivation presents the noiseless separable structure for clarity.

Before constructing the frequency-domain observation matrix, the two additional guard lines are discarded, and only the central \(M\) frequency lines are retained. To expose the underlying structure of the retained samples, we reorganize them into an $M\times(U_fN)$ matrix $\bm R'$ by writing $k=mU_fN+k'$ and defining $R'_{m,k'}\triangleq R_{\mathrm{FD}}[mU_fN+k']$, which can be written as
\begin{align}
R'_{m,k'}
&=
\sum_{p=1}^{P}
\underbrace{
e^{-j2\pi t_{d,p}\frac{m}{T}}
}_{E'_{m,p}}
\nonumber\\
&\quad\times
\underbrace{
\alpha_p
S\!\left(
\frac{k'}{U_fNT}-f_{D,p}
\right)
e^{-j2\pi t_{d,p}\frac{k'}{U_fNT}}
}_{V'_{p,k'}} .
\label{eq:Rprime_separable}
\end{align}
where $m=-\frac{M}{2},\ldots,\frac{M}{2}-1$ and $k'=-\frac{U_f N}{2},\ldots,\frac{U_f N}{2}-1$.
Accordingly, exploiting the separable structure in~\eqref{eq:Rprime_separable}, we define
$E'_{m,p}
\triangleq
e^{-j 2\pi t_{d,p} \frac{m}{T}},$
and $V'_{p,k'}
\triangleq
\alpha_p
S\!\left(
\frac{k'}{U_f N T} - f_{D,p}
\right)
e^{-j 2\pi t_{d,p} \frac{k'}{U_f N T}}.$
This allows \eqref{eq:Rprime_separable} to be expressed in matrix form as
\begin{equation}
\bm{R}' = \bm{E}' \bm{V}' .
\label{R'=E'V'}
\end{equation}

Starting from the observation matrix $\bm R'$, we form $T'_{k',q}=R'_{M/2-1-q,k'}$,
where $q=0,\ldots,M-1$ and $k'=-U_fN/2,\ldots,U_fN/2-1$. By setting \(\hat P_d=M-1\), the coefficient vector \(\bm a'_d\) is estimated from the approximate null space of \(\bm T'\), with its leading coefficient normalized to one. The roots of the resulting polynomial, denoted by \(Z'_{d,p}\), yield the delay estimates \(\hat t_{d,p}\) via~\eqref{hat_t}. With the delay estimates obtained, the matrices in \eqref{R'=E'V'} are reconstructed as
\begin{align}
\hat{\bm{E}}' &= \big( e^{-j 2\pi \hat t_{d,p} m/T} \big)_{m,p}, \\
\hat{\bm{V}}' &= ( \hat{\bm{E}}' )^\dagger \bm{R}', 
\end{align}
followed by compensation for the delay-dependent phase term,
\begin{align}
\widetilde V_{p,k'}'
=
\widehat V_{p,k'}'
e^{j2\pi\hat t_{d,p}\frac{k'}{U_fNT}}.
\end{align}
Applying a \(U_fN\)-point inverse DFT (IDFT) to \(\widetilde V_{p,k'}'\), with the constant normalization factor omitted, yields
\begin{align}
Y_p'[n]
=
\sum_{k'=-U_fN/2}^{U_fN/2-1}
\widetilde V_{p,k'}'
e^{j\frac{2\pi}{U_fN}nk'}.
\end{align}
Finally, the Doppler shifts $\hat f_{D,p}$ associated with $\hat t_{d,p}$ are recovered using the same Prony-based procedure as in the Doppler-first stage, now applied to $\bm{Y}'_p=(Y_p'[n])_{n=0}^{N-1}$ where only the central $N$ bins are retained. This completes the delay-first estimation procedure.

\subsection{Fusion of the Two Procedures} \label{merge}
The Doppler-first and delay-first estimators exhibit complementary failure modes: the former may fail when multiple paths exhibit similar Doppler shifts, while the latter may become unreliable when delays are closely spaced. To address these limitations, a fusion strategy is employed to jointly refine the path candidates obtained from both estimation orders.

Let $(\hat t_{d,p}^{(1)}, \hat f_{D,p}^{(1)})$, $p=1,\ldots,\hat P_D$, and
$(\hat t_{d,p}^{(2)}, \hat f_{D,p}^{(2)})$, $p=1,\ldots,\hat P_d$, denote the estimates obtained by the Doppler-first and delay-first methods, respectively.
For each branch, preliminary path gains are estimated by least squares, and candidates whose gain magnitudes are smaller than a pre-screening threshold \(\delta_\alpha\) times the maximum gain magnitude in the same branch are removed before fusion.

Let $\delta_t$ and $\delta_f $ denote the decision thresholds for delay and Doppler, respectively.
The delay and Doppler resolutions are defined as $\Delta t = T/M$ and $\Delta f = 1/(NT)$, respectively.
For any two estimates indexed by \((p,i)\) and \((q,j)\) with \((p,i)\neq(q,j)\), 
if \(\bigl| \hat t_{d,p}^{(i)}-\hat t_{d,q}^{(j)} \bigr|<\delta_t\Delta t\) and 
\(\bigl| \hat f_{D,p}^{(i)}-\hat f_{D,q}^{(j)} \bigr|<\delta_f\Delta f\), 
they are regarded as corresponding to the same physical path, with the delay and Doppler differences evaluated periodically over \(T\) and \(1/T\), respectively. Such candidates are merged sequentially, with the merged estimate updated by cumulative averaging after periodic alignment.

After fusion, a gain-based refinement is applied again to the fused candidates to further remove residual spurious path candidates. 
Let \(\Theta\) denote the fused set of delay--Doppler pairs, with cardinality \(\tilde P = |\Theta|\). 
Candidates whose refined gain magnitudes are smaller than a final pruning fraction \(\delta'_\alpha\) of the maximum refined gain are removed.
The remaining candidates are taken as the final detected paths, whose number is denoted by \(\hat P_{\mathrm{final}}\).

\section{Simulation Results}
\label{sect:simulation_results}

The performance of the proposed parallel Prony (PP) method is evaluated and compared with the single Doppler-first Prony (SP) scheme introduced in~\cite{YutakaPIMRC2025}. 
The adopted channel profile is a sparse physical multipath channel, where the path delays and Doppler shifts are independently generated from continuous uniform distributions. Since these continuous physical parameters are not constrained to the discrete delay--Doppler sampling lattice, the propagation paths are generally off-grid.
The simulation parameters are summarized in Table~\ref{tab:simulation_parameters}. 

\begin{table}[t]
\centering
\caption{Simulation parameters.}
\label{tab:simulation_parameters}
\footnotesize
\renewcommand{\arraystretch}{1.08}
\begin{tabular}{p{0.42\linewidth}|p{0.48\linewidth}}
\hline
Parameter & Value / Setting \\
\hline
Waveform & OTFS single-pilot waveform\\
DD grid size & \(N=M=32\) \\
Nominal transmit duration & \((N+2)T\) \\
Occupied bandwidth & \((M+2)/T\) \\
Delay / Doppler resolution & \(\Delta t=T/M\), \(\Delta f=1/(NT)\) \\
Oversampling factors & \(U_t=U_f=2\) \\
Tail parameter & \(N_0=2\) \\
Delay / Doppler distribution & \([0,T]\) / \([-1/(2T),\,1/(2T)]\), uniform \\
Path gains & Complex Gaussian, regenerated if \(|\alpha_p|<0.3\) \\
PP merging thresholds & \(\delta_t=0.1,\ \delta_f=0.5\) \\
Amplitude thresholds & Pre-screening: \(\delta_\alpha=0.01\), final pruning: \(\delta'_\alpha=0.1\)\\
Monte Carlo runs & \(1000\) \\
\hline
\end{tabular}
\end{table}

Additional explanations on the simulation settings are provided as follows.
Although \(N=M=32\) is a moderate-size setting, it is sufficient to evaluate fractional delay--Doppler mismatch and limited path separability, which are the main challenges addressed by the proposed method. 
The simulations are intended to evaluate the proposed off-grid physical delay--Doppler parameter estimator under the adopted OTFS pilot waveform.
For the sinc-type pilot waveform, the tail parameter \(N_0\) determines the additional observation interval used to reduce boundary truncation. In general, \(N_0\) should be selected according to the waveform tail energy, oversampling factors, and channel delay/Doppler spreads.
A too small \(N_0\) may degrade the DFT-based approximation, whereas an excessively large \(N_0\) introduces additional noise-dominated samples and computational cost. 
In the considered setting, \(N_0=2\) was empirically found to provide a sufficient observation margin, and this parameter can be adjusted according to the actual system configuration and channel conditions.

We first discuss the computational characteristics of the proposed PP method. Compared with the SP method, PP introduces additional computations mainly from the delay-first branch, oversampled observations, pairwise candidate fusion, pruning, and final least-squares gain refinement. Under $M=N$ with fixed oversampling factors, the least-squares matrix for the annihilating-filter coefficients has $O(N)$ rows and $O(N)$ columns, leading to an $O(N^3)$ dominant computational cost. The root finding of a degree-$(N-1)$ polynomial also has cubic-order complexity. Since the second-stage order is fixed to $L=1$, it does not change the leading complexity order. Thus, compared with a single fixed-order Doppler-first estimator, PP introduces a constant-factor increase in computational cost. On the other hand, by fixing the candidate orders, PP avoids the repeated order search over $\hat{P}=1,\ldots,N-1$ required in the SP method~\cite{YutakaPIMRC2025}. Therefore, the additional computational cost of PP remains moderate despite the extra delay-first branch and fusion steps.

For performance evaluation, the estimated and true paths are associated using the Hungarian algorithm. 
Since delay and Doppler are represented over finite normalized intervals, the association cost is computed using periodic distances. 
Specifically, the delay and Doppler periods are \(T\) and \(1/T\), respectively. 
A propagation path is declared successfully detected if both its estimated delay and Doppler deviate from the ground truth by less than half of the corresponding resolution bins, i.e., \(\Delta t/2\) and \(\Delta f/2\), after periodic wrapping. According to this criterion, the miss detection rate (MDR) and false alarm rate (FAR) are defined as
\begin{align}
\mathrm{MDR}
&=
1-\frac{1}{\mathrm{Runs}}
\sum_{r=1}^{\mathrm{Runs}}
\frac{\mathrm{CE}_r}{P},
\label{eq:MDR}\\
\mathrm{FAR}
&=
\frac{1}{\mathrm{Runs}}
\sum_{r=1}^{\mathrm{Runs}}
\frac{\hat P_{\mathrm{final},r}-\mathrm{CE}_r}
{\hat P_{\mathrm{final},r}},
\label{eq:FAR}
\end{align}
where \(P\) denotes the number of true propagation paths, \(\hat P_{\mathrm{final},r}\) is the total number of detected paths in the \(r\)-th run, and \(\mathrm{CE}_r\) represents the number of correctly estimated paths. 

\begin{figure}[t]
  \centering
  \includegraphics[
    width=1.0\linewidth,
    height=0.36\textheight,
    keepaspectratio
  ]{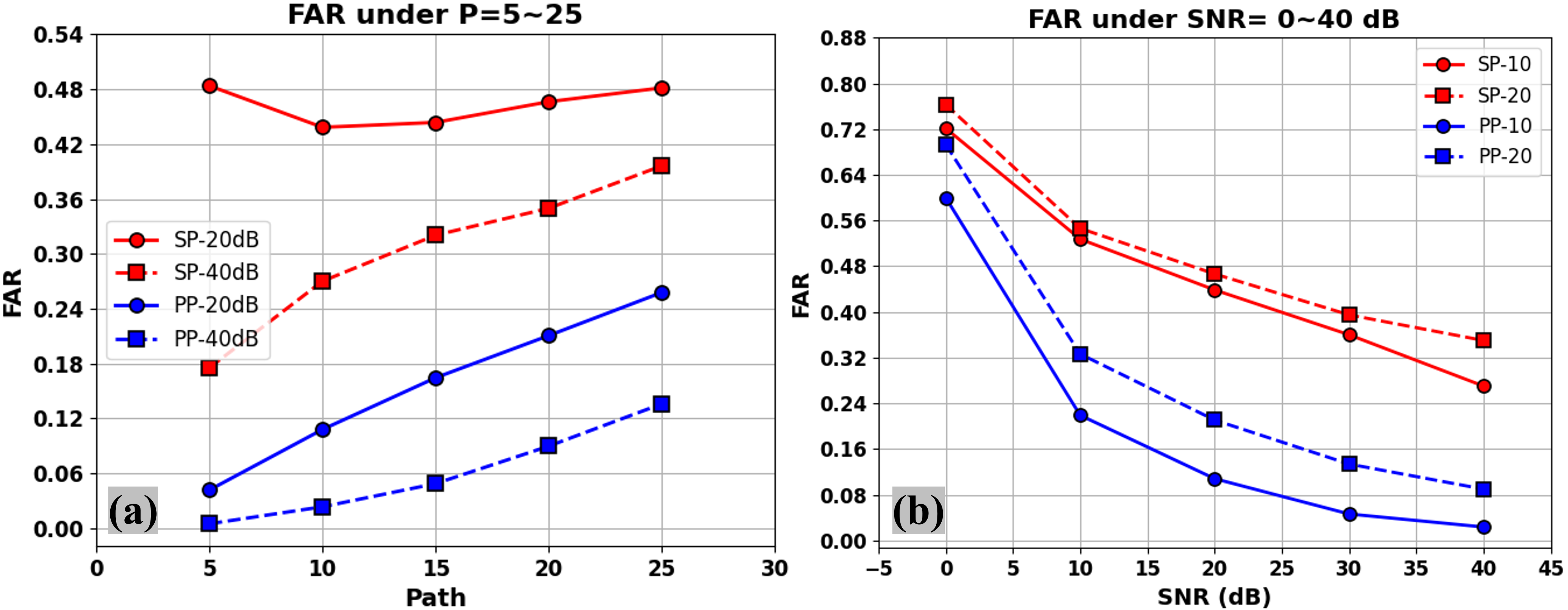}
  \capwidth=1.0\linewidth
    \caption{False alarm rate (FAR) performance comparison between SP and PP. 
    (a) FAR versus the number of paths under SNR = 20 and 40 dB. 
    (b) FAR versus SNR under \(P=10\) and \(P=20\).}
\label{compare_far}
\end{figure}

\begin{figure}[t]
  \centering
  \includegraphics[
    width=1.0\linewidth,
    height=0.36\textheight,
    keepaspectratio
  ]{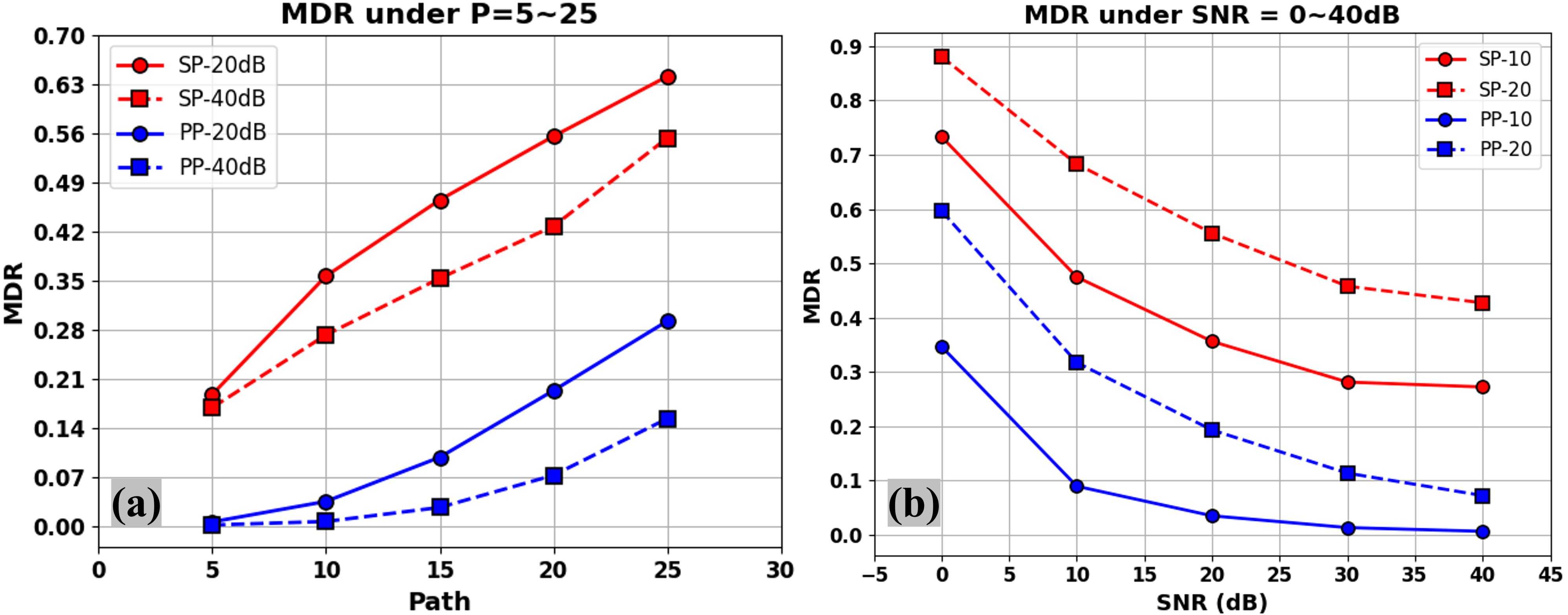}
    \caption{Miss detection rate (MDR) performance comparison between SP and PP. 
    (a) MDR versus the number of paths under SNR = 20 and 40 dB. 
    (b) MDR versus SNR under \(P=10\) and \(P=20\).}
\label{compare_mdr}
\end{figure}

Fig.~\ref{compare_far} and Fig.~\ref{compare_mdr} illustrate the FAR and MDR performance of both methods, respectively. 
As the number of propagation paths increases, the SP scheme exhibits rapid performance degradation, characterized by simultaneous increases in FAR and MDR under different SNR conditions. 
This behavior indicates that the single Doppler-first structure is vulnerable to inter-path interference and noise-induced spurious components in dense multipath environments. 
In contrast, the proposed PP method maintains lower FAR and MDR and experiences milder performance degradation as the number of paths increases, demonstrating improved robustness against multipath density. 
With increasing SNR, both methods benefit from reduced noise; however, PP provides more reliable detection performance, especially in low-to-moderate SNR regimes where spurious components and missed detections are more likely to occur.

Beyond detection performance, estimation accuracy is further quantified using the conditional root mean square error (cRMSE) of delay and Doppler. 
The cRMSE is computed from correctly matched paths only. 
Specifically, for the \(r\)-th run, let \(\mathrm{RMSE}_{\tau,\mathrm{correct}}^{(r)}\) and \(\mathrm{RMSE}_{f,\mathrm{correct}}^{(r)}\) denote the RMSEs of delay and Doppler over correctly matched paths, respectively. 
The corresponding cRMSEs are defined as
\begin{align}
\mathrm{cRMSE}_{\tau}
&=
\frac{1}{\mathrm{Runs}}
\sum_{r=1}^{\mathrm{Runs}}
\mathrm{RMSE}_{\tau,\mathrm{correct}}^{(r)},
\label{eq:cRMSE_delay}\\
\mathrm{cRMSE}_{f}
&=
\frac{1}{\mathrm{Runs}}
\sum_{r=1}^{\mathrm{Runs}}
\mathrm{RMSE}_{f,\mathrm{correct}}^{(r)}.
\label{eq:cRMSE_doppler}
\end{align}
The periodic delay and Doppler errors are used in the RMSE calculation.

\begin{figure}[t]
  \centering
  \includegraphics[
    width=1.0\linewidth,
    height=0.36\textheight,
    keepaspectratio
  ]{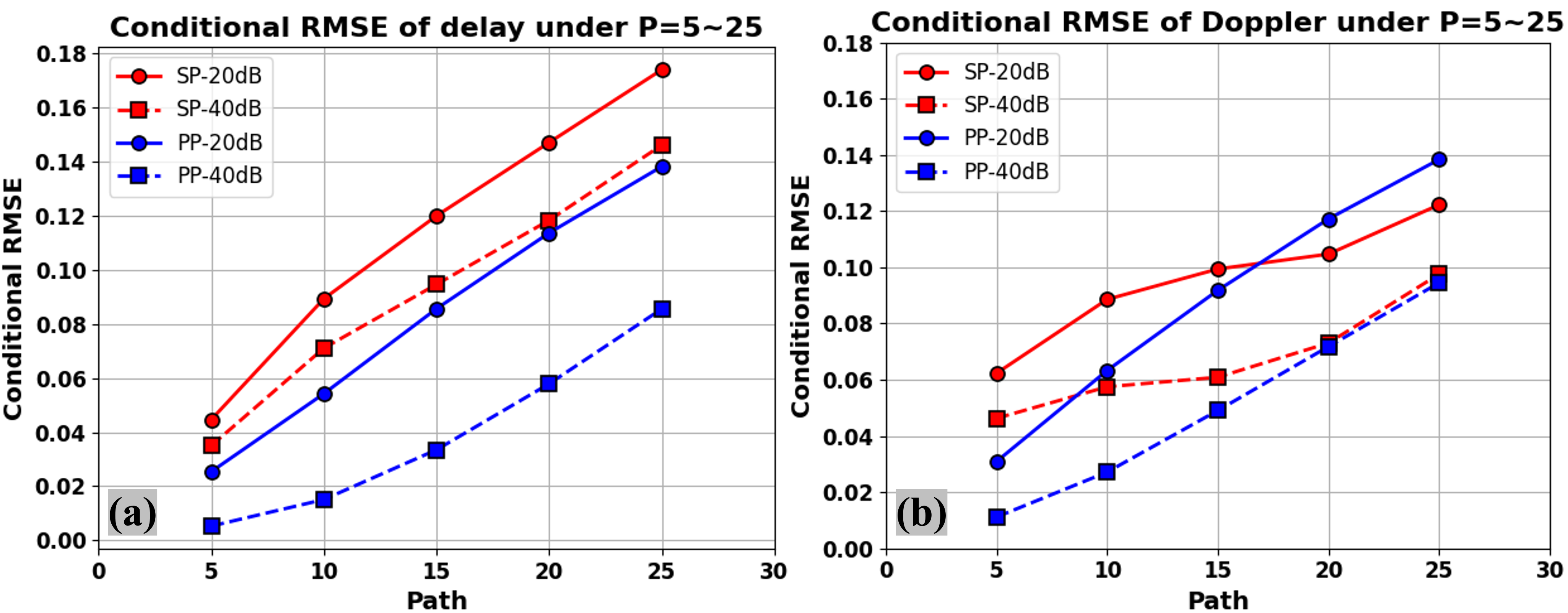}
  \capwidth=1.0\linewidth
    \caption{Conditional RMSE comparison between SP and PP versus the number of paths under SNR = 20 and 40 dB. 
    (a) Delay estimation. 
    (b) Doppler estimation.}
\label{compare_crmse_path}
\end{figure}

\begin{figure}[t]
  \centering
  \includegraphics[
    width=1.0\linewidth,
    height=0.36\textheight,
    keepaspectratio
  ]{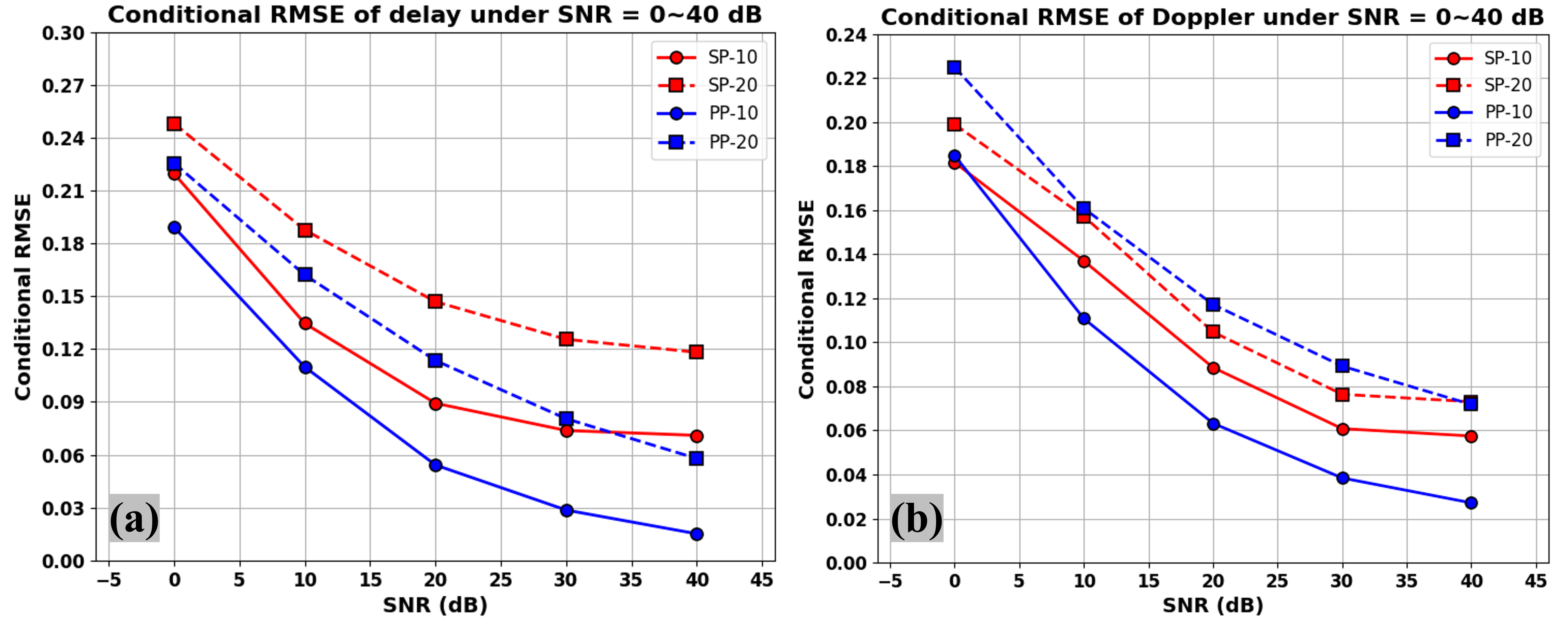}
  \capwidth=1.0\linewidth
    \caption{Conditional RMSE comparison between SP and PP versus SNR under \(P=10\) and \(P=20\). 
    (a) Delay estimation. 
    (b) Doppler estimation.}
\label{compare_crmse_snr}
\end{figure}

As shown in Fig.~\ref{compare_crmse_path} and Fig.~\ref{compare_crmse_snr}, the PP method generally achieves lower delay cRMSE than SP over a wide range of path numbers and SNRs, reflecting its improved robustness against noise and multipath density. 
However, for Doppler estimation, a different tendency can be observed in several cases. 
As shown in Fig.~\ref{compare_crmse_path}(b) and Fig.~\ref{compare_crmse_snr}(b), SP achieves comparable or slightly lower Doppler cRMSE than PP in some high-density cases, such as \(P=20\) and \(P=25\) at SNR \(=20\) dB.
One possible reason is the conditional nature of cRMSE, which is evaluated only over correctly detected paths.
Since SP tends to miss weak or closely spaced components, its Doppler cRMSE is often computed over relatively easier paths, as also reflected by the MDR and FAR results in Fig.~\ref{compare_far} and Fig.~\ref{compare_mdr}. 
Another factor is cross-stage error propagation in the PP framework: the delay-first branch relies on DFT-based processing, and residual delay estimation errors may affect the subsequent Doppler estimation. These errors may then be carried into the final estimates through the fusion process. 
Consequently, when SP correctly detects relatively well-separated or high-amplitude paths, it can sometimes achieve slightly lower Doppler cRMSE, even though its overall detection reliability is lower. Therefore, although the proposed PP framework improves path detectability and overall detection reliability, its Doppler estimation accuracy for difficult or closely spaced paths can still be further improved.

To further characterize typical estimation behavior while suppressing the influence of outliers, the median RMSE (mRMSE) of delay and Doppler is also evaluated. 
For the \(r\)-th run, let \(\mathrm{RMSE}_{\tau,\mathrm{all}}^{(r)}\) and \(\mathrm{RMSE}_{f,\mathrm{all}}^{(r)}\) denote the RMSEs computed using all matched path pairs obtained by the Hungarian assignment. 
The corresponding median RMSEs are defined as
\begin{align}
\mathrm{mRMSE}_{\tau}
&=
\operatorname{median}
\left\{
\mathrm{RMSE}_{\tau,\mathrm{all}}^{(r)}
\right\}_{r=1}^{\mathrm{Runs}},
\label{eq:mRMSE_delay}\\
\mathrm{mRMSE}_{f}
&=
\operatorname{median}
\left\{
\mathrm{RMSE}_{f,\mathrm{all}}^{(r)}
\right\}_{r=1}^{\mathrm{Runs}}.
\label{eq:mRMSE_doppler}
\end{align}
As with cRMSE, periodic delay and Doppler errors are used.

\begin{figure}[t]
  \centering
  \includegraphics[
    width=1.0\linewidth,
    height=0.36\textheight,
    keepaspectratio
  ]{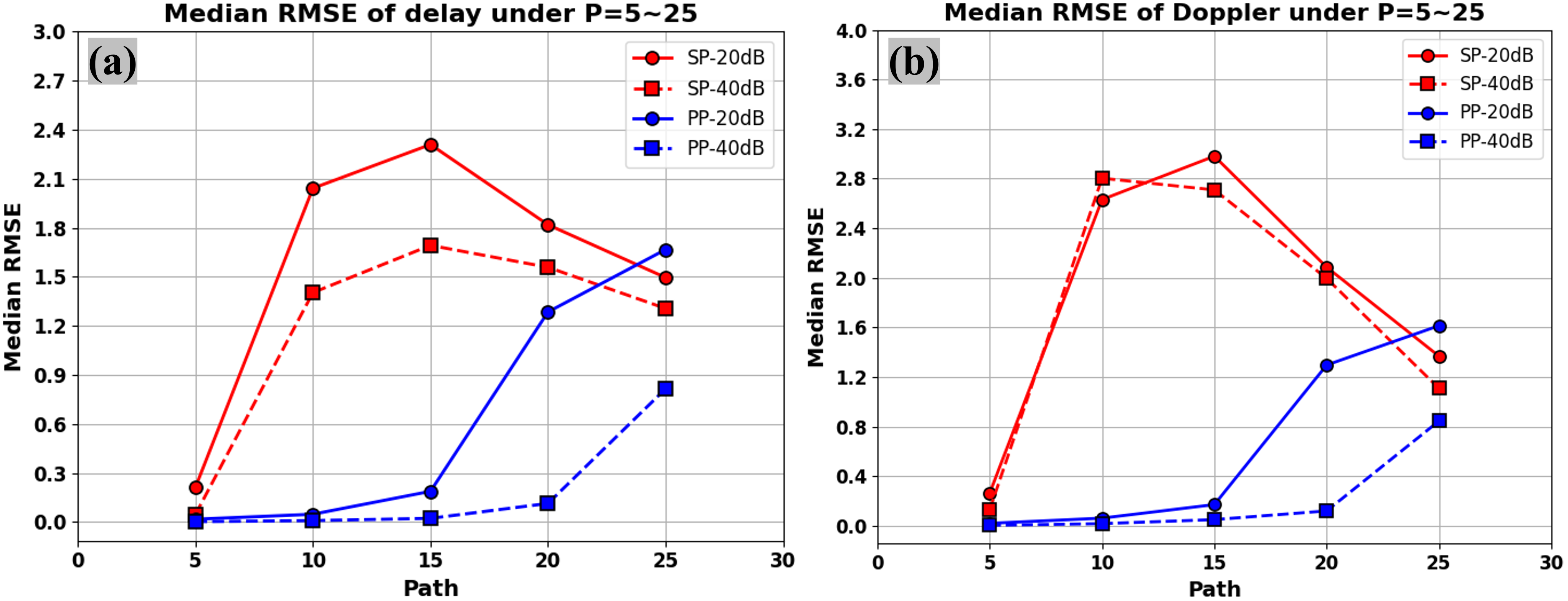}
  \capwidth=1.0\linewidth
    \caption{Median RMSE comparison between SP and PP versus the number of paths under SNR = 20 and 40 dB. 
    (a) Delay estimation. 
    (b) Doppler estimation.}
\label{compare_mrmse_path}
\end{figure}

\begin{figure}[t]
  \centering
  \includegraphics[
    width=1.0\linewidth,
    height=0.36\textheight,
    keepaspectratio
  ]{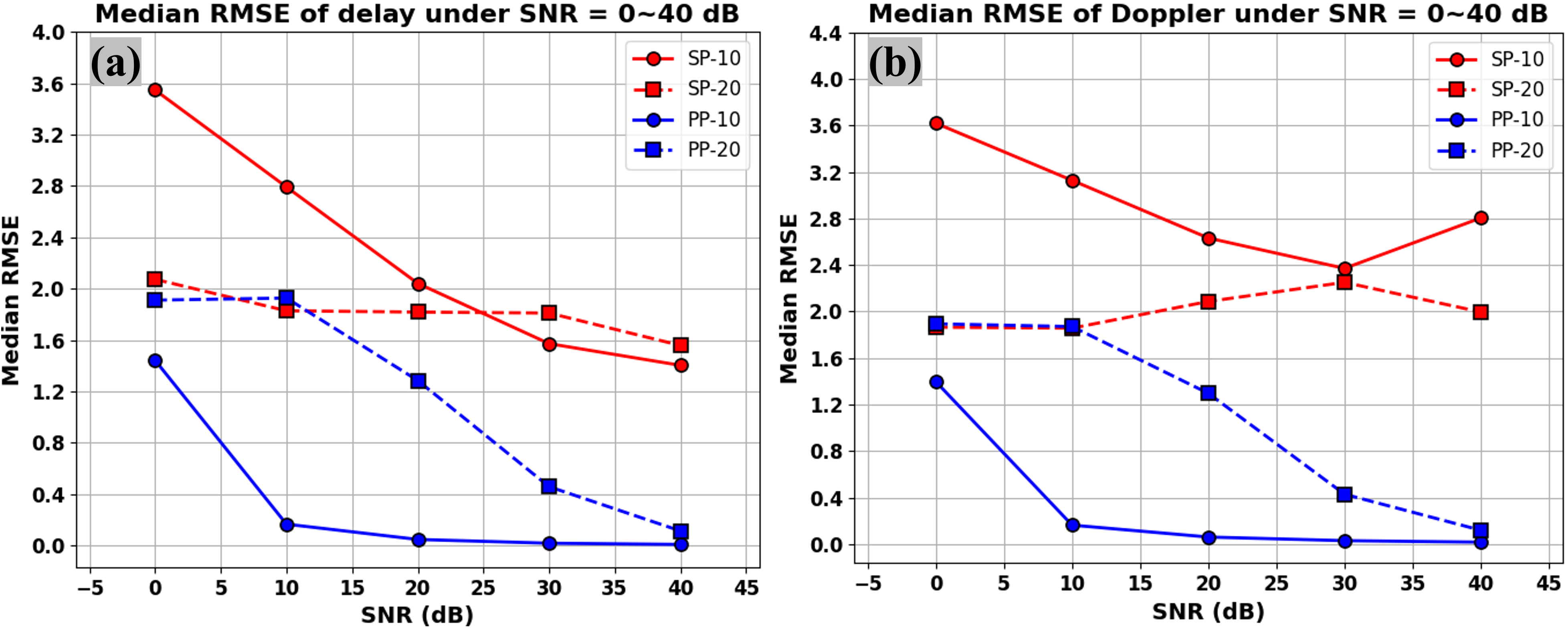}
  \capwidth=1.0\linewidth
    \caption{Median RMSE comparison between SP and PP versus SNR under \(P=10\) and \(P=20\). 
    (a) Delay estimation. 
    (b) Doppler estimation.}
\label{compare_mrmse_snr}
\end{figure}

As illustrated in Fig.~\ref{compare_mrmse_path} and Fig.~\ref{compare_mrmse_snr}, the proposed PP method generally achieves lower and more stable mRMSE than SP over most simulated path numbers and SNR values. 
This indicates that the parallel fusion of the Doppler-first and delay-first branches improves the stability of the matched-pair estimation results. 
However, some non-monotonic trends can also be observed for SP. 
On the one hand, the mRMSE does not always vary monotonically with \(P\). 
This may be because a larger \(P\) introduces more false candidates, which affect the Hungarian assignment and may reduce the matched-pair RMSE in some cases. 
On the other hand, in dense multipath cases, increasing SNR does not necessarily reduce the mRMSE clearly, since SP is not limited by noise alone but also by path separability. 
In particular, when several paths have similar Doppler shifts but different delays, the Doppler-first SP structure has inherent difficulty in separating them. 
These interpretations are consistent with Fig.~\ref{compare_far} and Fig.~\ref{compare_mdr}, where larger \(P\) results in higher MDR and FAR for SP. 
Therefore, the two RMSE metrics should be interpreted jointly with FAR and MDR.

\begin{table}[t]
\centering
\caption{Main structural and computational differences between SP and PP.}
\label{tab:sp_pp_comparison}
\footnotesize
\renewcommand{\arraystretch}{1.08}
\begin{tabular}{p{0.36\linewidth}|p{0.54\linewidth}}
\hline
Item & Difference \\
\hline
Estimation order 
& SP uses only the Doppler-first order, whereas PP performs Doppler-first and delay-first estimation in parallel. \\
\hline
Candidate order 
& SP searches over \(\hat P=1,\ldots,N-1\), whereas PP fixes \(\hat P_D=N-1\) and \(\hat P_d=M-1\). \\
\hline
Oversampling 
& PP employs time- and frequency-domain oversampling with \(U_t=U_f=2\) to improve the accuracy of the delay-first branch, whereas SP does not explicitly use this structure. \\
\hline
Fusion and pruning 
& PP performs pairwise candidate fusion and final gain-based pruning, which are not used in SP. \\
\hline
Additional cost 
& The main additional cost of PP comes from the delay-first branch, oversampled observations, pairwise fusion, and final least-squares gain estimation. \\
\hline
\end{tabular}
\end{table}

Overall, these results demonstrate that the proposed PP method improves detection reliability and estimation stability compared with the conventional single-order SP scheme, particularly in low-SNR and dense multipath scenarios. 

\section{Concluding Remarks}

This paper presented a Prony-based parallel two-stage framework for joint physical delay and Doppler estimation based on the single origin-pilot waveform adopted in this work over doubly selective physical channels.
The adopted waveform has a frequency-shifted sinc pulse structure and differs from conventional rectangular-pulse OTFS implementations. 
Instead of relying on a simplified on-grid equivalent delay--Doppler channel representation, the received signal model was formulated from sampled observations of the known pilot waveform, enabling direct estimation of continuous physical delay and Doppler parameters for sensing-oriented applications.
By combining Doppler-first and delay-first estimation branches and fusing their outputs, the proposed PP method exploits complementary delay--Doppler-domain information to improve robustness against off-grid mismatch, noise, and inter-path interference. 
Numerical results showed that PP improves detection reliability and estimation stability compared with the conventional single-order SP scheme, especially in low-SNR and dense multipath environments. 
These results indicate that the proposed framework is a promising approach for high-precision off-grid physical delay--Doppler parameter estimation in high-mobility sensing and future ISAC applications.

\appendix
\section{Comparison of Four Delay--Doppler Waveform Realizations and Their Observation Models}\label{sec:model}

\subsection{General Condition}
\label{subsec:tx}

We consider four representative continuous-time transmit realizations to examine how different DD-domain modulation schemes construct their transmit waveforms, particularly in terms of pulse shaping and windowing. These include ISFFT-OTFS, originally proposed by Hadani et al.~\cite{OTFS}; Type-1 Zak-OTFS and Type-2 Zak-OTFS, two implementation classes characterized by Gopalam et al.~\cite{gopalam2024zak}; and ODDM, introduced by Lin and Yuan~\cite{Lin_ODDM}. These realizations represent the conventional ISFFT-based construction, two Zak-based signal-generation procedures, and DD-domain orthogonal-pulse modulation, respectively. Recent comparative studies, such as in~\cite{gopalam2024zak, shafie2025spectrum}, have also investigated the waveform characteristics and differences among these realizations.

Let $X_{\rm DD}[k,l]$ denote the DD-domain symbols, where $l=0,\ldots,M-1$ and $k=0,\ldots,N-1$ are the delay and Doppler indices, respectively. The corresponding DD coordinates are
\begin{equation}
\tau_l=\frac{lT}{M},
\qquad
\nu_k=\frac{k}{NT},
\label{eq:app_dd_coordinates}
\end{equation}
where $T$ is the time-slot duration and the frame contains $N$ time slots.

\subsubsection{General waveform}

\paragraph{ISFFT-OTFS}
The conventional ISFFT-OTFS realization follows the original OTFS principle of representing information symbols in the DD domain and spreading them across the TF grid to exploit channel diversity in both time and frequency~\cite{OTFS}. This mapping is based on the symplectic finite Fourier transform (SFFT), a variant of the two-dimensional discrete Fourier transform with opposite signs in the exponential terms along the two dimensions. Under the DFT normalization convention adopted in this work, the forward SFFT is unnormalized, while the inverse SFFT carries a factor of $1/(MN)$. Accordingly, the SFFT maps the TF-domain symbols to the DD domain as
\begin{equation}
\begin{aligned}
X_{\rm DD}[k,l]
&=
\sum_{m=0}^{M-1}
\sum_{n=0}^{N-1}
X_{\rm TF}[m,n]\\
&\quad\times
e^{-j2\pi(nk/N-ml/M)}.
\end{aligned}
\label{eq:sfft}
\end{equation}
where $m=0,\ldots,M-1$ and $n=0,\ldots,N-1$ denote the frequency and time-slot indices, respectively.

For OTFS transmission, the inverse SFFT (ISFFT) is applied to the DD-domain symbols to obtain the corresponding TF-domain symbols:
\begin{equation}
\begin{aligned}
X_{\rm TF}[m,n]
&=
\frac{1}{MN}
\sum_{l=0}^{M-1}
\sum_{k=0}^{N-1}
X_{\rm DD}[k,l]\\
&\quad\times
e^{j2\pi(nk/N-ml/M)}.
\end{aligned}
\label{eq:isfft}
\end{equation}
The continuous-time transmitted signal is then generated by Heisenberg synthesis with the subcarrier spacing $\Delta f=1/T$~\cite{OTFS}:
\begin{equation}
x_{\rm I}(t)
=
\sum_{n=0}^{N-1}
\sum_{m=0}^{M-1}
X_{\rm TF}[m,n]\,
g_{\rm tx}(t-nT)
e^{j2\pi m\Delta f(t-nT)},
\label{eq:heisenberg}
\end{equation}
where $g_{\rm tx}(t)$ denotes the transmit pulse.

Substituting \eqref{eq:isfft} into \eqref{eq:heisenberg} and regrouping the terms associated with each DD symbol gives
\begin{align}
x_{\rm I}(t)
&=
\frac{1}{MN}
\sum_{l=0}^{M-1}
\sum_{k=0}^{N-1}
X_{\rm DD}[k,l]
\sum_{n=0}^{N-1}
e^{j2\pi nk/N}
\nonumber\\
&\qquad\times
g_{\rm tx}(t-nT)
\sum_{m=0}^{M-1}
e^{j2\pi m[\Delta f(t-nT)-l/M]}
\nonumber\\
&\overset{(1)}{=}
\frac{1}{MN}
\sum_{l=0}^{M-1}
\sum_{k=0}^{N-1}
X_{\rm DD}[k,l]\,
D_M\!\left(\frac{t-\tau_l}{T}\right)
\nonumber\\
&\qquad\times
\sum_{n=0}^{N-1}
e^{j2\pi nk/N}
g_{\rm tx}(t-nT),
\label{eq:isfft_substitution}
\end{align}
where
\begin{equation}
D_M(u)
=
\sum_{m=0}^{M-1}e^{j2\pi mu}
\label{eq:dirichlet}
\end{equation}
denotes the finite Dirichlet kernel. Equality~(1) follows from $e^{-j2\pi mn}=1$ for integer $m$ and $n$.

\paragraph{Type-1 Zak-OTFS}

The Zak transform provides a delay-Doppler representation of a time-domain signal by applying a Fourier transform across successive time intervals. For a continuous-time signal $u(t)$, the Zak transform with period $T$ is defined as~\cite{gopalam2024zak}
\begin{equation}
\mathcal{Z}_T\{u\}(\tau,\nu)
=
\sum_{q=-\infty}^{\infty}
u(\tau+qT)e^{-j2\pi\nu qT},
\label{eq:zak_transform}
\end{equation}
where $\tau$ and $\nu$ denote the delay and Doppler variables, respectively. Unlike conventional ISFFT-OTFS, which explicitly maps DD-domain symbols to the TF domain before waveform generation, the discrete Zak transform enables direct mapping between the DD domain and discrete-time sequences, eliminating the intermediate TF-domain representation~\cite{lampel2022discrete}.

Gopalam et al.~\cite{gopalam2024zak} further developed two Zak-OTFS realizations, termed Type-1 and Type-2, based on different orders of time-domain windowing and frequency-domain pulse shaping. Type-1 applies time-domain windowing before pulse shaping, whereas Type-2 performs pulse shaping before time-domain windowing. Since these operations generally do not commute, the two realizations produce different continuous-time waveforms. In the corresponding DD-domain twisted-convolution formulations, the Type-2 filter contains an additional phase factor $e^{j2\pi\tau\nu}$ relative to Type-1, as can be seen by comparing Eqs.~(21) and~(22) of~\cite{gopalam2024zak}.

We first consider the Type-1 Zak-OTFS realization. Under the inverse-DFT normalization convention adopted in~\eqref{eq:idzt}, the delay-time coefficients are obtained by an $N$-point IDFT along the Doppler dimension~\cite{lampel2022discrete,gopalam2024zak}:
\begin{equation}
X_{\rm TD}[l,n]
=
\frac{1}{N}
\sum_{k=0}^{N-1}
X_{\rm DD}[k,l]e^{j2\pi nk/N},
\label{eq:app_idzt}
\end{equation}
where $l=0,\ldots,M-1$ and $n=0,\ldots,N-1$.

Let $w(t)$ denote a frame-level time-domain window supported on $[0,NT)$, and denote its sampled
values by
\begin{equation}
w[q]
\triangleq
w\!\left(\frac{qT}{M}\right),
\qquad q\in\mathbb Z.
\label{eq:app_window_samples}
\end{equation}
The coefficient $X_{\rm TD}[l,n]$ corresponds to the time instant
\begin{equation}
\frac{(nM+l)T}{M}=nT+\tau_l.
\end{equation}
Time-domain windowing therefore gives
\begin{equation}
\widetilde X_{\rm TD}[l,n]
=
w[nM+l]X_{\rm TD}[l,n].
\label{eq:z1_windowed_coefficients}
\end{equation}

Let $p(t)$ denote the elementary pulse used for continuous-time pulse shaping. Each windowed coefficient weights a copy of $p(t)$ centered at $nT+\tau_l$, yielding
\begin{equation}
\begin{aligned}
x_{\rm Z1}(t)
&=
\sum_{n=0}^{N-1}
\sum_{l=0}^{M-1}
\widetilde X_{\rm TD}[l,n]\,
p(t-\tau_l-nT)\\
&=
\sum_{n=0}^{N-1}
\sum_{l=0}^{M-1}
w[nM+l]X_{\rm TD}[l,n]\,
p(t-\tau_l-nT).
\end{aligned}
\label{eq:z1_synthesis}
\end{equation}

Substituting \eqref{eq:app_idzt} into \eqref{eq:z1_synthesis} and regrouping the terms associated with each DD symbol gives
\begin{align}
x_{\rm Z1}(t)
&=
\frac{1}{N}
\sum_{n=0}^{N-1}
\sum_{l=0}^{M-1}
\sum_{k=0}^{N-1}
X_{\rm DD}[k,l]e^{j2\pi nk/N}
\nonumber\\
&\qquad\times
w[nM+l]p(t-\tau_l-nT)
\nonumber\\
&=
\frac{1}{N}
\sum_{l=0}^{M-1}
\sum_{k=0}^{N-1}
X_{\rm DD}[k,l]
\nonumber\\
&\qquad\times
\sum_{n=0}^{N-1}
w[nM+l]e^{j2\pi nk/N}
p(t-\tau_l-nT).
\label{eq:z1_substitution}
\end{align}

\paragraph{Type-2 Zak-OTFS}

The Type-2 Zak-OTFS realization uses the same IDZT but applies pulse shaping before time-domain windowing~\cite{gopalam2024zak}. For this construction, the delay-time coefficients are periodically extended along $n$:
\begin{equation}
X_{\rm TD}[l,n+N]=X_{\rm TD}[l,n],
\qquad n\in\mathbb Z.
\label{eq:app_td_periodic_extension}
\end{equation}
Accordingly, the expression in \eqref{eq:app_idzt} holds for all integer $n$.

Before time-domain windowing, pulse shaping gives
\begin{equation}
x_P(t)
=
\sum_{n\in\mathbb Z}
\sum_{l=0}^{M-1}
X_{\rm TD}[l,n]\,
p(t-\tau_l-nT).
\label{eq:z2_pulse_shaping}
\end{equation}
The continuous-time frame window is then applied to the pulse-shaped signal:
\begin{align}
x_{\rm Z2}(t)
&=
w(t)x_P(t)
\nonumber\\
&=
w(t)
\sum_{n\in\mathbb Z}
\sum_{l=0}^{M-1}
X_{\rm TD}[l,n]\,
p(t-\tau_l-nT).
\label{eq:z2_synthesis}
\end{align}

Substituting \eqref{eq:app_idzt} into \eqref{eq:z2_synthesis} and regrouping the terms associated with each DD symbol gives
\begin{align}
x_{\rm Z2}(t)
&=
\frac{w(t)}{N}
\sum_{n\in\mathbb Z}
\sum_{l=0}^{M-1}
\sum_{k=0}^{N-1}
X_{\rm DD}[k,l]e^{j2\pi nk/N}
\nonumber\\
&\qquad\times
p(t-\tau_l-nT)
\nonumber\\
&=
\frac{w(t)}{N}
\sum_{l=0}^{M-1}
\sum_{k=0}^{N-1}
X_{\rm DD}[k,l]
\nonumber\\
&\qquad\times
\sum_{n\in\mathbb Z}
e^{j2\pi nk/N}
p(t-\tau_l-nT).
\label{eq:z2_substitution}
\end{align}

\paragraph{ODDM}

While Zak-OTFS eliminates the intermediate TF-domain mapping through the inverse Zak transform, ODDM provides a more direct approach to waveform synthesis by modulating DD-domain symbols without explicitly performing either the ISFFT or the IDZT.

Proposed by Lin and Yuan~\cite{Lin_ODDM}, ODDM directly constructs the continuous-time waveform through delay shifts and Doppler modulation of a DD-domain orthogonal pulse (DDOP). The DDOP is formed from a finite train of elementary pulses and is designed to achieve local orthogonality at the fine delay and Doppler resolutions of the DD grid.

While the original ODDM design employs a square-root raised-cosine (SRRC) elementary pulse~\cite{Lin_ODDM}, we use the same general elementary pulse $p(t)$ as in the Zak-OTFS descriptions for the following structural comparison. In~\ref{subsec:sr_specialization}, we specialize $p(t)$ to the frequency-shifted sinc pulse $p_{\rm T}(t)$ defined in~\eqref{eq:pulse_time}. With this common pulse, the DDOP prototype is defined as
\begin{equation}
u(t)
=
\frac{1}{N}
\sum_{n=0}^{N-1}p(t-nT).
\label{eq:app_ddop}
\end{equation}
Here, the factor $1/N$ sets the pulse-train scaling consistently with the inverse-DFT normalization adopted in the two Zak realizations.

The transmitted signal is obtained by applying
the delay shift $\tau_l$ and the continuous-time
Doppler modulation $\nu_k$ to $u(t)$:
\begin{equation}
x_{\rm O}(t)
=
\sum_{l=0}^{M-1}
\sum_{k=0}^{N-1}
X_{\rm DD}[k,l]\,
u(t-\tau_l)
e^{j2\pi\nu_k(t-\tau_l)}.
\label{eq:oddm_synthesis}
\end{equation}
Substituting \eqref{eq:app_ddop} into
\eqref{eq:oddm_synthesis} gives
\begin{align}
x_{\rm O}(t)
&=
\frac{1}{N}
\sum_{l=0}^{M-1}
\sum_{k=0}^{N-1}
X_{\rm DD}[k,l]\,
e^{j2\pi\nu_k(t-\tau_l)}
\nonumber\\
&\qquad\times
\sum_{n=0}^{N-1}
p(t-\tau_l-nT).
\label{eq:oddm_substitution}
\end{align}

\subsubsection{Comparison of the Four Continuous-Time Transmit Realizations}
\label{subsec:waveform_comparison}
We compare the four realizations in terms of windowing and pulse shaping, the actions of the delay and Doppler coordinates, temporal and spectral support, and orthogonality requirements. The elementary pulse and its Fourier transform are denoted by $p(t)$ and $P(f)$, respectively, while $w(t)$ and $W(f)$ denote the time-domain window and its Fourier transform. ODDM refers to the direct continuous-time synthesis defined above. For each realization $q\in\{\mathrm I,\mathrm{Z1},\mathrm{Z2},\mathrm O\}$, let $x^{(q)}_{l,k}(t)$ denote the waveform multiplying $X_{\rm DD}[k,l]$ in the preceding transmit expression. The transmitted signal for arbitrary DD symbols can therefore be written as
\begin{equation}
x_q(t)
=
\sum_{l=0}^{M-1}
\sum_{k=0}^{N-1}
X_{\rm DD}[k,l]\,x^{(q)}_{l,k}(t).
\label{eq:comparison_dd_expansion}
\end{equation}

\paragraph{Windowing Objects and Pulse-Shaping Order}

The first distinction concerns whether the window acts on discrete coefficients or a continuous-time waveform, and whether it is applied at the slot or frame level. Table~\ref{tab:tx_windowing_comparison} summarizes these differences for the four realizations considered here~\cite{OTFS,gopalam2024zak,Lin_ODDM,shafie2025spectrum}.

\begin{table*}[t]
\centering
\caption{Windowing and waveform synthesis in the four transmit realizations.}
\label{tab:tx_windowing_comparison}
\renewcommand{\arraystretch}{1.15}
\setlength{\tabcolsep}{5pt}
\begin{tabular}{p{0.15\textwidth}p{0.38\textwidth}p{0.38\textwidth}}
\hline
Realization & Waveform synthesis & Windowing object \\
\hline
ISFFT-OTFS
& DD-to-TF precoding followed by slotwise multicarrier synthesis. & $g_{\rm tx}(t-nT)$ acts on the multicarrier component associated with slot $n$. \\
Type-1 Zak-OTFS
& IDZT, sample windowing, and interpolation using the common pulse $p(t)$.
& $w[nM+l]$ weights the discrete TD coefficients before interpolation. \\
Type-2 Zak-OTFS
& Pulse shaping of the periodically extended TD sequence, followed
by continuous-time windowing.
& $w(t)$ acts on the complete pulse-shaped waveform. \\
ODDM
& Construction of a finite DDOP, followed by delay shifts and
continuous-time Doppler modulation.
& The time-domain envelope is incorporated into $u(t)$; no separate
fixed frame window is applied in the direct synthesis considered here. \\
\hline
\end{tabular}
\end{table*}

Type-1 and Type-2 Zak-OTFS differ in the order of windowing and pulse shaping, which leads to different treatments of pulse contributions around the frame boundaries. In Type-1, only the windowed discrete coefficients contribute to pulse shaping, although the resulting pulse tails may extend beyond the nominal frame and overlap with adjacent blocks. Similarly, ODDM constructs a finite DDOP from elementary pulses and retains their tails in the direct continuous-time synthesis~\cite{gopalam2024zak,Lin_ODDM}. 

In contrast, Type-2 performs pulse shaping on the periodically extended sequence before applying a fixed frame window. Consequently, pulse contributions originating outside the frame can affect the waveform within the frame, while the final output is confined to the support of the window. 

For rectangular-slot ISFFT-OTFS, boundary processing instead occurs at each time slot. Such blockwise restriction can introduce boundary discontinuities, whereas linear pulse shaping preserves the associated transients and allows pulse tails to overlap across adjacent blocks~\cite{bayat2025unified}. Truncating these tails to enforce a finite transmission interval modifies the spectrum and may introduce additional out-of-band leakage ~\cite{gopalam2024zak}. 

These differences reflect the tradeoff between temporal occupancy and spectral confinement.

\paragraph{Action of the Delay Coordinate}

As shown in \eqref{eq:isfft_substitution} and \eqref{eq:z2_substitution}, the DD delay coordinate shifts the waveform-generating components in both ISFFT-OTFS and Type-2 Zak-OTFS, while their corresponding windows remain fixed. This conclusion concerns the DD encoding index rather than the physical propagation delay, which acts on the complete transmitted waveform. In ISFFT-OTFS, $\tau_l$ shifts the Dirichlet kernel $D_M((t-\tau_l)/T)$, whereas the slot windows $g_{\rm tx}(t-nT)$ remain fixed. In Type-2 Zak-OTFS, $\tau_l$ shifts the infinite pulse train through $p(t-\tau_l-nT)$, whereas the frame window $w(t)$ remains fixed.

Still, the two realizations also differ in where the window-induced mismatch occurs.
In rectangular-slot ISFFT-OTFS, each multicarrier symbol is windowed separately, so the mismatch can occur at every time-slot boundary. In Type-2 Zak-OTFS, the window acts on the entire pulse-shaped waveform, so under rectangular windowing, the mismatch occurs at the frame boundaries, where the shifted infinite pulse train is truncated by the fixed window. For a general nonconstant frame window, differences in window weighting may also occur within the frame. 
Nevertheless, despite these differences, both realizations share the same overall property: changing the DD delay index does not generally translate the complete transmitted waveform~\cite{OTFS,gopalam2024zak}.

As shown in \eqref{eq:z1_substitution} and \eqref{eq:oddm_substitution}, both Type-1 Zak-OTFS and ODDM introduce the DD delay coordinate through the shifted elementary pulses $p(t-\tau_l-nT)$. Unlike ISFFT and Type-2, these pulse trains are not subsequently truncated by fixed slot or frame windows, thereby avoiding the window-boundary mismatch described above.

In ODDM, the delay coordinate shifts the entire fixed prototype $u(t)$ together with its continuous-time Doppler modulation. Therefore, the complete waveform satisfies
\begin{equation}
x^{(\mathrm O)}_{l,k}(t)
=
x^{(\mathrm O)}_{0,k}(t-\tau_l).
\label{eq:comparison_oddm_delay_shift}
\end{equation}
In Type-1, however, the pulse weights $w[nM+l]$ also depend on the delay index $l$. Thus, changing $l$ may alter both the pulse positions and their amplitudes. Comparing \eqref{eq:z1_substitution} with the corresponding waveform at $l=0$, the same complete-waveform delay translation holds if
\begin{equation}
w[nM+l]=w[nM]
\label{eq:comparison_z1_window_condition}
\end{equation}
for every contributing $n$. Under this condition, Type-1 and ODDM share the same complete-waveform delay translation property. In particular, the rectangular sample window considered in the following subsection satisfies this condition.

\paragraph{Location of Doppler Modulation}

Similar to the delay coordinate, the Doppler coordinate also allows the four realizations to be broadly divided into two groups.
As shown in \eqref{eq:isfft_substitution} and \eqref{eq:z1_substitution}, both ISFFT-OTFS and Type-1 Zak-OTFS introduce the DD Doppler coordinate through the discrete phase coefficients $e^{j2\pi\nu_knT}$. In ISFFT, these coefficients weight the multicarrier components associated with individual time slots, whereas in Type-1, they weight the individual shifted elementary pulses. Thus, in both realizations, Doppler modulation is incorporated through phase weights assigned to discrete slot or pulse indices, rather than through an explicitly applied continuous-time modulation factor.

In contrast, as shown in \eqref{eq:z2_substitution} and \eqref{eq:oddm_substitution}, Type-2 Zak-OTFS and ODDM represent the Doppler coordinate through the continuous-time factor $e^{j2\pi\nu_k(t-\tau_l)}$, but their underlying constructions differ. In Type-2, the infinite pulse train admits an exact decomposition into a continuous-time Doppler factor and a periodic kernel:
\begin{equation}
x^{(\mathrm{Z2})}_{l,k}(t)
=
\frac{w(t)}{N}
e^{j2\pi\nu_k(t-\tau_l)}
Q_k(t-\tau_l),
\label{eq:comparison_z2_periodic_form}
\end{equation}
where
\begin{align}
Q_k(v)
&=
\frac1T
\sum_{m\in\mathbb Z}
P\!\left(\nu_k+\frac mT\right)
e^{j2\pi mv/T},
\nonumber\\
Q_k(v+T)&=Q_k(v).
\label{eq:comparison_z2_periodic_kernel}
\end{align}
The periodic kernel is determined by the frequency samples of $P(f)$ and generally depends on the Doppler index $k$~\cite{gopalam2024zak}.

In ODDM, the continuous-time Doppler factor is instead applied directly to the same finite prototype $u(t)$ for every Doppler index. Compared with Type-1, this introduces an additional intra-pulse phase factor $e^{j2\pi\nu_k(t-\tau_l-nT)}$ beyond the discrete phase coefficient $e^{j2\pi\nu_knT}$. Therefore, although Type-2 and ODDM both exhibit continuous-time Doppler modulation, the former uses a $k$-dependent periodic kernel, whereas the latter modulates a fixed finite prototype~\cite{Lin_ODDM,shafie2025spectrum}. Overall, these distinctions concern how Doppler modulation is incorporated into the respective synthesis expressions, rather than whether the phase of the resulting waveform is continuous or discontinuous.

\paragraph{Spectral Characteristics}

The four realizations also exhibit different spectral structures due to their distinct waveform synthesis procedures. For Type-1 Zak-OTFS and ODDM, the key distinction is whether the elementary pulse spectrum shifts with the Doppler index. Taking the Fourier transforms of their component waveforms gives
\begin{align}
\widehat{x}^{(\mathrm{Z1})}_{l,k}(f)
&=
\frac{e^{-j2\pi f\tau_l}}{N}P(f)
\nonumber\\
&\quad\times
\sum_{n=0}^{N-1}w[nM+l]e^{-j2\pi(f-\nu_k)nT},
\label{eq:comparison_z1_spectrum}\\
\widehat{x}^{(\mathrm O)}_{l,k}(f)
&=
\frac{e^{-j2\pi f\tau_l}}{N}P(f-\nu_k)
\nonumber\\
&\quad\times
\sum_{n=0}^{N-1}e^{-j2\pi(f-\nu_k)nT}.
\label{eq:comparison_oddm_spectrum}
\end{align}
Type-1 uses the common multiplicative spectral envelope $P(f)$ across Doppler indices, whereas ODDM uses the Doppler-shifted envelope $P(f-\nu_k)$. Thus, Type-1 changes the interference pattern within a common envelope, while ODDM also shifts the envelope itself~\cite{gopalam2024zak,Lin_ODDM,shafie2025spectrum}.

For Type-2 Zak-OTFS, the elementary pulse spectrum determines the weights of discrete frequency components rather than appearing as a continuous multiplicative envelope. Defining $f_{m,k}=\nu_k+m/T$ and transforming \eqref{eq:comparison_z2_periodic_form} gives
\begin{align}
\widehat{x}^{(\mathrm{Z2})}_{l,k}(f)
&=
\frac1{TN}\sum_{m\in\mathbb Z}
P(f_{m,k})e^{-j2\pi f_{m,k}\tau_l}
\nonumber\\
&\quad\times W(f-f_{m,k}).
\label{eq:comparison_z2_spectrum}
\end{align}
Here, $P(f_{m,k})$ determines the frequency-component weights, while $W(f-f_{m,k})$ describes the spectral broadening of each component due to frame-level windowing~\cite{gopalam2024zak}.

In contrast, the ISFFT-OTFS subcarrier components involve $G_{\rm tx}(f-m/T)$, where $G_{\rm tx}(f)=\mathcal F\{g_{\rm tx}(t)\}$. Their spectral shapes are therefore determined by the slotwise transmit pulse rather than a frame-level window~\cite{OTFS}.

\paragraph{Orthogonality Requirements}

Orthogonality requires the inner product between any two distinct DD-indexed basis waveforms to vanish, so that the corresponding symbols can be separated without mutual interference under ideal correlation processing. Specifically, for a given realization and any two distinct DD indices $(l,k)\ne(l',k')$, the orthogonality condition is
\begin{equation}
\left\langle x_{l,k},x_{l',k'}\right\rangle
=
\int_{\mathbb R}
x_{l,k}^*(t)x_{l',k'}(t)\,dt
=0.
\label{eq:comparison_general_orthogonality}
\end{equation}
When this condition holds, correlating the transmitted waveform with a basis waveform extracts the corresponding DD symbol, multiplied by the energy of that basis waveform, without interference from the other symbols.

The four realizations establish orthogonality of their complete basis waveforms through different constructions. In ISFFT-OTFS, the transmitted waveform is generated from TF-domain pulses with time and frequency spacings $T$ and $1/T$. If these TF pulses are mutually orthogonal, the scaled ISFFT preserves their mutual orthogonality~\cite{OTFS,lampel2022discrete}. In ODDM, orthogonality is imposed directly on the finite prototype $u(t)$, whose ambiguity function must satisfy
\begin{equation}
\mathcal A_{u,u}
\left(
\frac{\Delta l T}{M},
\frac{\Delta k}{NT}
\right)=0
\label{eq:comparison_oddm_orthogonality}
\end{equation}
for nonzero index differences within the prescribed DD grid~\cite{Lin_ODDM}. 
Here, $\mathcal A_{g,h}(\tau,\nu)$ denotes the cross-ambiguity function of two signals $g(t)$ and $h(t)$, defined as
\begin{equation*}
\mathcal A_{g,h}(\tau,\nu)
=
\int_{\mathbb R}
g\!\left(t+\frac{\tau}{2}\right)
h^*\!\left(t-\frac{\tau}{2}\right)
e^{-j2\pi\nu t}\,dt.
\end{equation*}

For the two Zak-OTFS realizations, the IDZT alone does not guarantee orthogonality after pulse shaping and windowing. Their respective pulse and window requirements can be derived from the inner products of their complete DD-indexed basis waveforms.

For Type-1 Zak-OTFS, the orthogonality of delayed elementary pulses can be examined through the pulse autocorrelation
\begin{equation}
R_p(\Delta)
=
\int_{\mathbb R}p^*(t)p(t+\Delta)\,dt.
\end{equation}
Consider two pulses shifted to different delay-grid positions, $p_l(t)=p(t-lT/M)$ and $p_{l'}(t)=p(t-l'T/M)$. Their inner product is
\begin{equation}
\left\langle p_l,p_{l'}\right\rangle
=
R_p\left(\frac{(l-l')T}{M}\right).
\end{equation}
Therefore, to ensure that elementary pulses at distinct integer delay-grid positions are mutually orthogonal, a sufficient pulse-design condition is the Nyquist shift-orthogonality condition
\begin{equation}
R_p\left(\frac{qT}{M}\right)
=
E_p\delta_{q,0},
\qquad q\in\mathbb Z,
\label{eq:comparison_zak_nyquist}
\end{equation}
where $E_p=R_p(0)$ is the pulse energy, and $\delta_{q,0}$ is the Kronecker delta, equal to one when $q=0$ and zero otherwise. This condition makes the inner product vanish for distinct integer delay-grid shifts.

To examine the orthogonality of the complete DD-indexed basis waveforms, substituting the Type-1 component waveforms into their inner product gives
\begin{equation}
\begin{aligned}
&\left\langle
x^{(\mathrm{Z1})}_{l,k},
x^{(\mathrm{Z1})}_{l',k'}
\right\rangle\\
&=
\frac1{N^2}
\sum_{n=0}^{N-1}
\sum_{n'=0}^{N-1}
w^*[nM+l]w[n'M+l']\\
&\quad\times e^{j2\pi(k'n'-kn)/N}\\
&\quad\times
R_p(\tau_l+nT-\tau_{l'}-n'T).
\end{aligned}
\label{eq:comparison_z1_general_gram}
\end{equation}
The argument of the autocorrelation can be expressed as
\begin{equation}
\tau_l+nT-\tau_{l'}-n'T
=
\frac{(n-n')M+l-l'}{M}T.
\end{equation}
Since $0\le l,l'<M$, the integer $(n-n')M+l-l'$ is zero only when $n=n'$ and $l=l'$. Therefore, imposing the Nyquist condition in~\eqref{eq:comparison_zak_nyquist} on $p(t)$ eliminates all terms except those with $n=n'$ and $l=l'$, yielding
\begin{equation}
\begin{aligned}
&\left\langle
x^{(\mathrm{Z1})}_{l,k},
x^{(\mathrm{Z1})}_{l',k'}
\right\rangle\\
&=
\frac{E_p\delta_{l,l'}}{N^2}
\sum_{n=0}^{N-1}
|w[nM+l]|^2
e^{j2\pi(k'-k)n/N}.
\end{aligned}
\label{eq:comparison_z1_orthogonality}
\end{equation}
Thus, the Nyquist condition on $p(t)$ ensures orthogonality between different delay indices. For each fixed $l$, the remaining sum is a DFT coefficient of the window-energy sequence $|w[nM+l]|^2$. To ensure orthogonality between different Doppler indices, all its nonzero DFT coefficients must vanish. Since the DFT is invertible, this holds if and only if
\begin{equation}
|w[nM+l]|=c_l,
\qquad n=0,\ldots,N-1,
\end{equation}
where $c_l$ is a positive constant that may depend on $l$. Hence, the discrete window must have constant magnitude across time slots for each fixed delay index, while its phase is unrestricted. Under these pulse and window conditions,
\begin{equation}
\begin{aligned}
&\left\langle
x^{(\mathrm{Z1})}_{l,k},
x^{(\mathrm{Z1})}_{l',k'}
\right\rangle
=
\frac{E_pc_l^2}{N}\delta_{l,l'}\delta_{k,k'}.
\end{aligned}
\label{eq:comparison_z1_orth_result}
\end{equation}

For Type-2 Zak-OTFS, the continuous-time window is applied after pulse shaping~\cite{gopalam2024zak}. Substituting the component waveforms into their inner product gives
\begin{equation}
\begin{aligned}
&\left\langle
x^{(\mathrm{Z2})}_{l,k},
x^{(\mathrm{Z2})}_{l',k'}
\right\rangle\\
&=
\frac{
e^{j2\pi(\nu_k\tau_l-\nu_{k'}\tau_{l'})}
}{N^2}
\int_{\mathbb R}
|w(t)|^2
e^{j2\pi(\nu_{k'}-\nu_k)t}\\
&\quad\times
Q_k^*(t-\tau_l)
Q_{k'}(t-\tau_{l'})\,dt.
\end{aligned}
\label{eq:comparison_z2_gram}
\end{equation}
Unlike Type-1, the window weight remains inside the continuous-time integral. To preserve orthogonality between the frequency lines of $Q_k$, we impose a constant-magnitude frame window satisfying $|w(t)|=c_w>0$ for $0\le t<NT$ and $w(t)=0$ outside the frame. Since $\nu_{k'}-\nu_k=(k'-k)/(NT)$, the frequency lines satisfy
\begin{equation}
\begin{aligned}
&\int_0^{NT}
e^{j2\pi[(\nu_{k'}-\nu_k)+(m'-m)/T]t}\,dt\\
&\qquad=NT\delta_{k,k'}\delta_{m,m'}.
\end{aligned}
\end{equation}
Consequently, all terms with different frequency lines vanish, and the inner product reduces to
\begin{equation}
\begin{aligned}
&\left\langle
x^{(\mathrm{Z2})}_{l,k},
x^{(\mathrm{Z2})}_{l',k'}
\right\rangle\\
&=
\frac{c_w^2}{NT}\delta_{k,k'}
e^{j2\pi\nu_k(\tau_l-\tau_{l'})}\\
&\quad\times
\sum_{m\in\mathbb Z}
\left|P\left(\nu_k+\frac{m}{T}\right)\right|^2
e^{j2\pi m(l-l')/M}.
\end{aligned}
\label{eq:comparison_z2_flat_gram}
\end{equation}
Thus, the window condition ensures orthogonality between different Doppler indices.

The remaining requirement on $p(t)$ can be derived through a DFT argument analogous to that used for the Type-1 window. Writing $m=r+qM$, the spectral sum becomes
\begin{equation}
\begin{aligned}
&\sum_{m\in\mathbb Z}
\left|P\left(\nu_k+\frac{m}{T}\right)\right|^2
e^{j2\pi m(l-l')/M}\\
&=
\sum_{r=0}^{M-1}
e^{j2\pi r(l-l')/M}
\sum_{q\in\mathbb Z}
\left|P\left(\nu_k+\frac{r+qM}{T}\right)\right|^2.
\end{aligned}
\end{equation}
For each fixed $k$, this expression is an $M$-point DFT of the accumulated spectral energies across the $M$ frequency-index classes. Analogously to the Type-1 window condition, orthogonality between different delay indices requires all the nonzero DFT coefficients to vanish. By the invertibility of the DFT, this holds if and only if
\begin{equation}
\sum_{q\in\mathbb Z}
\left|
P\left(\nu_k+\frac{r+qM}{T}\right)
\right|^2
=C_k>0,
\quad r=0,\ldots,M-1,
\label{eq:comparison_z2_spectral_condition}
\end{equation}
where $C_k$ is independent of $r$. Therefore, for each fixed Doppler index $k$, the pulse spectrum must have equal accumulated energies across the $M$ frequency-index classes. Under these pulse and window conditions,
\begin{equation}
\begin{aligned}
&\left\langle
x^{(\mathrm{Z2})}_{l,k},
x^{(\mathrm{Z2})}_{l',k'}
\right\rangle
=
\frac{Mc_w^2C_k}{NT}
\delta_{l,l'}\delta_{k,k'}.
\end{aligned}
\label{eq:comparison_z2_orth_result}
\end{equation}

The two Zak realizations therefore exhibit complementary DFT-based orthogonality structures: Type-1 uses the Nyquist pulse condition to establish delay orthogonality and equal discrete window energies to establish Doppler orthogonality, whereas Type-2 uses the frame-window condition to establish Doppler orthogonality and equal accumulated pulse-spectrum energies to establish delay orthogonality.

In summary, although all four realizations encode information symbols in the DD domain, they do not necessarily produce identical continuous-time transmit waveforms due to differences in how DD symbols are mapped into continuous-time signals. Specifically, they differ in how Doppler modulation is introduced, through discrete phase weighting or continuous-time modulation, and how time localization is implemented, through slot-level windowing, sample-level weighting, frame-level windowing, or finite-prototype construction. These structural differences lead to distinct delay-translation properties, spectral characteristics, and orthogonality requirements.

\subsection{Sinc-Pulse and Rectangular-Window Specialization}
\label{subsec:sr_specialization}

In this work, we adopt frequency-shifted sinc pulse shaping and rectangular windowing. Substituting these choices into the transmit expressions of the four realizations discussed in the preceding section allows us to further simplify their waveform structures and consolidate the distinctions among them. We retain the DD coordinates $\tau_l=lT/M$ and $\nu_k=k/(NT)$, with $l=0,\ldots,M-1$ and $k=0,\ldots,N-1$, and the amplitude factors used above. Let $M$ be even and $N\geq2$.

Unlike the two Zak-OTFS realizations and ODDM, conventional ISFFT-OTFS employs a rectangular time-slot pulse, given by $g_{\rm tx}(t)=1$ for $0\leq t<T$ and zero otherwise~\cite{OTFS}. Substituting this rectangular slot pulse into \eqref{eq:isfft_substitution}, only the term $n=\lfloor t/T\rfloor$ contributes within the frame. Retaining the Dirichlet kernel with carrier indices
$m=0,\ldots,M-1$, define
\begin{align}
d_M(v)&=D_M(v/T)
\nonumber\\
&=e^{j\pi(M-1)v/T}
\frac{\sin(\pi Mv/T)}{\sin(\pi v/T)}.
\label{eq:sr_dirichlet}
\end{align}
The transmitted waveform therefore simplifies to
\begin{align}
x_{\rm I}(t)
&=\frac{1}{MN}
\sum_{l=0}^{M-1}\sum_{k=0}^{N-1}
X_{\rm DD}[k,l]
\nonumber\\
&\quad\times
\underbrace{
e^{j2\pi k\lfloor t/T\rfloor/N}
}_{\substack{\text{discrete phase}\\
\text{weighting}}}
\nonumber\\
&\quad\times
\underbrace{
d_M(t-\tau_l)
}_{\substack{\text{Dirichlet-kernel}\\
\text{delay shift}}},
\quad 0\leq t<NT.
\label{eq:sr_isfft}
\end{align}
Outside this interval, $x_{\rm I}(t)=0$. The delay coordinate shifts the Dirichlet kernel, whereas the slotwise phase factor is determined by the fixed time-slot index $\lfloor t/T\rfloor$.

For Type-1 Zak-OTFS, the rectangular window gives $w[nM+l]=1$ for $n=0,\ldots,N-1$
and $l=0,\ldots,M-1$. Substituting this window and the frequency-shifted sinc pulse in \eqref{eq:pulse_time} into \eqref{eq:z1_substitution}, the transmitted waveform simplifies to
\begin{align}
x_{\rm Z1}(t)
&=\sum_{l=0}^{M-1}\sum_{k=0}^{N-1}
X_{\rm DD}[k,l]
\nonumber\\
&\quad\times
\underbrace{
x^{(\mathrm{Z1})}_{0,k}(t-\tau_l)
}_{\substack{\text{complete-waveform}\\
\text{delay translation}}},
\label{eq:sr_z1_operator}
\end{align}
where
\begin{align}
x^{(\mathrm{Z1})}_{0,k}(v)
&=\frac{M}{TN}
\sum_{n=0}^{N-1}
\underbrace{
e^{j2\pi nk/N}
}_{\substack{\text{discrete phase}\\
\text{weighting}}}
\nonumber\\
&\quad\times
\operatorname{sinc}\!\left(
\frac{M}{T}(v-nT)
\right)
e^{-j\pi(v-nT)/T}.
\label{eq:sr_z1_basis}
\end{align}
Here, $\sinc(x)$ denotes the normalized sinc function, defined as $\sinc(x)=\sin(\pi x)/(\pi x)$ for $x\neq 0$, with $\sinc(0)=1$.
The rectangular sample window is represented by the finite summation over $n$, without an additional frame window multiplying the resulting waveform.

Under the origin-pilot configuration $X_{\rm DD}[0,0]=1$, with all other DD
symbols set to zero, this expression reduces to
\begin{align}
x_{\rm Z1}(t)
&=\frac{1}{N}
\sum_{n=0}^{N-1}p_{\rm T}(t-nT)
 =s(t).
\label{eq:sr_z1_main_connection}
\end{align}
This coincides with the transmitted pilot waveform obtained from \eqref{eq:transmitted_signal} and \eqref{pilot-DDdomain}, including its normalization. The Dirichlet representation in \eqref{dirichlet_waveform} is subsequently obtained under the periodic approximation adopted in the main text.

For Type-2 Zak-OTFS, substituting the frequency-shifted sinc pulse in \eqref{eq:pulse_time} and the rectangular frame window into \eqref{eq:z2_substitution}, the transmitted waveform simplifies to
\begin{align}
x_{\rm Z2}(t)
&=\frac{1}{N}
\sum_{l=0}^{M-1}\sum_{k=0}^{N-1}
X_{\rm DD}[k,l]
\nonumber\\
&\quad\times
\underbrace{
e^{j2\pi\nu_k(t-\tau_l)}
}_{\substack{\text{continuous-time}\\
\text{Doppler modulation}}}
\nonumber\\
&\quad\times
\underbrace{
Q_k(t-\tau_l)
}_{\substack{\text{periodic-kernel}\\
\text{delay translation}}},
\quad 0\leq t<NT.
\label{eq:sr_z2_operator}
\end{align}
Here, $Q_k(v)$ is the periodic kernel defined in \eqref{eq:comparison_z2_periodic_kernel}, evaluated using the frequency-shifted sinc
spectrum in \eqref{eq:pulse_frequency}. Outside this interval, $x_{\rm Z2}(t)=0$.
The delay coordinate shifts the periodic kernel and its continuous modulation factor, whereas the rectangular frame boundaries remain fixed.

For ODDM, substituting the frequency-shifted sinc pulse in \eqref{eq:pulse_time} into the DDOP prototype in \eqref{eq:app_ddop}, we obtain
\begin{align}
x_{\rm O}(t)
&=\sum_{l=0}^{M-1}\sum_{k=0}^{N-1}
X_{\rm DD}[k,l]
\nonumber\\
&\quad\times
\underbrace{
x^{(\mathrm O)}_{0,k}(t-\tau_l)
}_{\substack{\text{complete-waveform}\\
\text{delay translation}}},
\label{eq:sr_oddm_operator}
\end{align}
where
\begin{equation}
x^{(\mathrm O)}_{0,k}(v)
=
\underbrace{
e^{j2\pi\nu_kv}
}_{\substack{\text{continuous-time}\\
\text{Doppler modulation}}}
u(v).
\label{eq:sr_oddm_basis}
\end{equation}
Both the finite pulse train and its continuous-time Doppler modulation depend on the shifted time argument $t-\tau_l$. No additional frame window is applied to the synthesized waveform.

Under the pulse and window settings specified above, \eqref{eq:sr_isfft}, \eqref{eq:sr_z1_operator}, \eqref{eq:sr_z2_operator}, and \eqref{eq:sr_oddm_operator} group the four realizations along two axes: complete-waveform delay translation and Doppler-modulation placement. Combining these two groupings gives the $2\times2$ comparison in Table~\ref{tab:sr_two_by_two}.

\begin{table*}[t]
\centering
\caption{Delay and Doppler actions under the adopted pulse and window settings.}
\label{tab:sr_two_by_two}
\begingroup
\renewcommand{\arraystretch}{1.2}
\setlength{\tabcolsep}{5pt}
\begin{tabular}{
p{0.28\textwidth}
p{0.32\textwidth}
p{0.32\textwidth}}
\hline
Delay action
& Discrete phase weighting
& Continuous-time Doppler modulation \\
\hline
Fixed slot/frame boundaries
& \textbf{ISFFT-OTFS}
& \textbf{Type-2 Zak-OTFS} \\
Complete-waveform delay translation
& \textbf{Type-1 Zak-OTFS}
& \textbf{ODDM} \\
\hline
\end{tabular}
\endgroup
\end{table*}

For the delay rows, as established in \eqref{eq:comparison_oddm_delay_shift} and
\eqref{eq:comparison_z1_window_condition}, ODDM satisfies complete-waveform delay translation, while Type-1 shares this property when $w[nM+l]=w[nM]$. The rectangular sample window considered here satisfies $w[nM+l]=w[nM]=1$ over the retained indices. Consequently, both realizations satisfy
\begin{equation}
x^{(q)}_{l,k}(t)=x^{(q)}_{0,k}(t-\tau_l),
\qquad q\in\{\mathrm{Z1},\mathrm O\}.
\label{eq:sr_complete_delay_shift}
\end{equation}
Under these settings, their difference lies only in the Doppler modulation: as shown in \eqref{eq:z1_substitution} and \eqref{eq:oddm_substitution}, ODDM introduces an additional intra-pulse phase factor $e^{j2\pi\nu_k(t-\tau_l-nT)}$ relative to Type-1. Both realizations retain the pulse tails without additional frame truncation. In contrast, ISFFT and Type-2 retain fixed slot or frame boundaries, so the complete-waveform translation identity does not generally hold. Similarly, this comparison concerns the DD encoding coordinate; a physical channel delay shifts the complete transmitted waveform in all four realizations.

For the Doppler columns, ISFFT and Type-1 apply the discrete phase weights $e^{j2\pi nk/N}$ to individual slots and pulses, respectively. Type-2 and ODDM instead exhibit the continuous-time factor $e^{j2\pi\nu_k(t-\tau_l)}$, multiplying the periodic kernel $Q_k$ obtained from the infinite pulse train and the finite prototype $u$, respectively. The Type-2 kernel $Q_k$ may itself depend on $k$.
Thus, the columns describe Doppler placement in the displayed synthesis expressions, not whether the phase of the resulting waveform is continuous.

Under the origin-pilot configuration adopted in this work, $X_{\rm DD}[0,0]=1$ and all other DD symbols are zero. Both the discrete Doppler weights and the continuous-time Doppler factors then become unity. Under this configuration, the additional intra-pulse phase factor $e^{j2\pi\nu_k(t-\tau_l-nT)}$ becomes unity, eliminating the remaining difference between Type-1 and ODDM. With the same frequency-shifted sinc pulse, both realizations therefore reduce to
\begin{align}
x_{\rm Z1}(t)
&=x_{\rm O}(t)=u(t)
\nonumber\\
&=\frac{1}{N}
\sum_{n=0}^{N-1}p_{\rm T}(t-nT)
\nonumber\\
&=s(t).
\label{eq:sr_origin_finite_train}
\end{align}
Thus, the two realizations yield exactly the same finite pulse train, coinciding with the main-text transmitted pilot including its normalization. Consistent with the waveform generation used in the simulations, as described in Section~\ref{sect:channel_model}, two additional temporal guard pulses and one guard frequency line on each side are introduced to mitigate boundary effects. Let \(p_{{\rm T},\mathrm g}(t)\) denote the corresponding frequency-shifted sinc pulse with an occupied bandwidth of \((M+2)/T\), given by
\begin{equation}
p_{{\rm T},\mathrm g}(t)
=
\frac{\sin\!\left(\pi(M+2)t/T\right)}{\pi t}
e^{-j\pi t/T}.
\end{equation}
Under the same origin-pilot configuration, extending the pulse train to \(N+2\) pulses in both realizations yields
\begin{align}
x_{\rm Z1}^{(\mathrm g)}(t)
&=x_{\rm O}^{(\mathrm g)}(t)
\nonumber\\
&=\frac{1}{N}
\sum_{n=0}^{N+1}
p_{{\rm T},\mathrm g}(t-nT)
\nonumber\\
&=s_{\mathrm g}(t),
\label{eq:sr_origin_guard_equivalence}
\end{align}
where \(s_{\mathrm g}(t)\) denotes the guard-extended transmitted pilot waveform used in the simulations. Thus, the two realizations remain exactly equivalent under the same guard extensions, with the normalization factor \(1/N\) unchanged.

This equivalence holds for the origin pilot only; for nonzero Doppler indices, e.g., when data symbols or a pilot with \(k\neq0\) are present, the two realizations differ as discussed above. We therefore refer to the adopted waveform as Type-1 Zak-OTFS in the main text.

For ISFFT and Type-2, the origin-pilot waveforms instead reduce to
\begin{align}
x_{\rm I}(t)
&=\frac{1}{MN}d_M(t),
\nonumber\\
x_{\rm Z2}(t)
&=\frac{e^{-j\pi Mt/T}}{TN}d_M(t)
\nonumber\\
&=\frac{M}{T}
e^{-j\pi Mt/T}x_{\rm I}(t),
\label{eq:sr_origin_periodic_pair}
\end{align}
for $0\leq t<NT$, with both waveforms zero outside this interval. They therefore share the same frame-windowed periodic structure, differing only by the displayed normalization and fixed frequency shift associated with their frequency-index conventions.

Consequently, under the origin-pilot configuration, the $2\times2$ comparison in
Table~\ref{tab:sr_two_by_two} reduces to two classes: the finite sinc-pulse train of Type-1 and ODDM, which retains the pulse tails without additional frame truncation, and the frame-windowed periodic Dirichlet waveform of ISFFT and Type-2.

\section{Receiver Filtering and Sampling}
\label{app:sampling}
The receiver front-end is an ideal LPF with passband $|f| < 1/(2T_s) = U_t M/(2T)$, followed by sampling at $T_s = T/(U_t M)$~\cite{oppenheim2010discrete}. The guard-extended pilot occupies $[-(M+3)/(2T),\,(M+1)/(2T))$. Since $|f_{D,p}| < 1/(2T)$, the received useful signal lies within $|f| < (M+4)/(2T)$.
Hence, it passes the LPF without distortion if $(M+4)/(2T) < U_t M/(2T)$, i.e., $M > 4/(U_t - 1)$ for $U_t > 1$, which holds for $U_t = 2$ and $M = 32$. The samples in (8) then coincide exactly with the continuous-time point samples. For AWGN with power spectral density $N_0$, the sampled noise $z_{\mathrm{TD}}[i]$ is white with variance $N_0/T_s$.



\bibliographystyle{ieicetr}
\bibliography{reference}



\end{document}